\documentclass[aps,pre,twocolumn,notitlepage,nofootinbib,superscriptaddress]{revtex4}
\newif\ifpdf\ifx\pdfoutput\undefined\pdffalse\else\pdfoutput=1\pdftrue\fi

\usepackage{graphicx}
\usepackage{color}
\newcommand{\avg}[1]{\langle{#1}\rangle}

\newcommand{\abs}[1]{\left|{#1}\right|}

\newcommand{\beq}{\begin{equation}}
\newcommand{\eeq}{\end{equation}} 

\newcommand{\beqar}{\begin{eqnarray}}
\newcommand{\eeqar}{\end{eqnarray}} 
\newcommand{\ignore}[1]{}
\newcommand{\comment}[1]{}
\newcommand{\nignore}[1]{}
\newcommand{\vect}[1]{{\bf #1}}
\newcommand{\mat}[1]{{\bf #1}}
\newcommand{\mr}[1]{{\mathrm{#1}}}

\newcommand{\Tm}{{\bf M}}
\newcommand{\tm}{{\bf m}}
\newcommand{\transp}{^{\rm T}}
\newcommand{\tG}{t_{\rm G}}
\newcommand{\Eort}{\mat{E}_{\rm C}}
\newcommand{\Eorti}{\mat{E}_{\rm C,0}}
\newcommand{\Eg}{\mat{E}_{\rm G}}
\newcommand{\gene}[2]{{\fcolorbox{black}{fun#1}{\makebox[30pt]{#2$_{}^{}$}}}}
\definecolor{yel}{rgb}{1,1,0}
\newcommand{\moduleannotation}[1]{\mbox{\textcolor{yel}{#1}}}
\newcommand{\vspac}[1]{\begin{picture}(0,#1)(0,0)\end{picture}}
\newcommand{\genesize}{\tiny}
\newcommand{\modulesize}{\normalsize}
\newcommand{\moduletitlesize}{\large}
\newcommand{\module}[1]{\setlength{\unitlength}{0.6pt}\fbox{\begin{minipage}{3.200in}\begin{flushleft}#1\end{flushleft}\end{minipage}}}

\definecolor{fun0}{rgb}{1,1,1}
\definecolor{black}{rgb}{0,0,0}
\definecolor{fun1}{rgb}{0.5,1,0.5}
\definecolor{fun2}{rgb}{0.5,0.5,1}
\definecolor{fun3}{rgb}{0.8,0.6,0.8}
\definecolor{fun4}{rgb}{1,0.5,0.6}

\begin{document}
\title{Finding regulatory modules through large-scale gene-expression data analysis} 
\author{Morten Kloster}
\affiliation{Department of Physics, Princeton University, Princeton, New Jersey 08544}
\affiliation{NEC Laboratories America, Inc., 4 Independence Way, Princeton, New Jersey 08540}
\author{Chao Tang}
\email[To whom correspondence should be adressed. E-mail:]{tang@nec-labs.com}
\affiliation{NEC Laboratories America, Inc., 4 Independence Way, Princeton, New Jersey 08540}
\affiliation{Center for Theoretical Biology, Peking University, Beijing 100871, China}
\author{Ned Wingreen}
\affiliation{NEC Laboratories America, Inc., 4 Independence Way, Princeton, New Jersey 08540}

\date{\today}

\begin{abstract}
\end{abstract}


\begin{abstract}
The use of gene microchips has enabled a rapid accumulation 
of gene-expression data.
One of the major challenges of analyzing this data is the diversity,
in both size and signal strength, of
the various modules in the gene regulatory networks of organisms. Based on
the Iterative Signature
Algorithm [Bergmann, S., Ihmels, J. \& Barkai, N. (2002) {\it Phys. Rev.
E} {\bf 67}, 031902],
we present an algorithm---the Progressive Iterative Signature Algorithm
(PISA)---that, by sequentially eliminating
modules, allows unsupervised identification of both large and 
small regulatory modules. We applied PISA to a large set of
yeast gene-expression data, and, using the Gene Ontology database
as a reference, found that the algorithm is much better able to 
identify regulatory modules than methods based on
high-throughput transcription-factor binding experiments or 
on comparative genomics.
\end{abstract}

\maketitle

\section{Introduction}

The introduction of DNA microarray technology has made it possible to
aquire vast 
amounts of gene expression data, raising the issue of how best to
extract information from this data.
While basic clustering algorithms have been successful at finding
genes 
that are coregulated for a small, specific set of experimental
conditions
~\cite{Alon_Broad_patterns,Eisen_Cluster_analysis,Tamayo_Interpreting_patterns},
these algorithms are less effective when applied to large data sets due
to two
well-recognized limitations. First, standard clustering algorithms
assign each gene to a single cluster, while many genes in fact 
belong to multiple transcriptional 
regulons.~\cite{Bittner_Data_analysis,Cheng_Biclustering_of,Gasch_Exploring_the,SA}.
Second, each transcriptional regulon may only be active in a few
experiments, and the remaining experiments will only contribute to the
noise~\cite{Getz_Coupled_two-way,Cheng_Biclustering_of,SA}.

A number of approaches have been proposed to overcome one or both of
these problems
\cite{Getz_Coupled_two-way,Califano_Analysis_of,Cheng_Biclustering_of,Owen_Gene_Recommender,Gasch_Exploring_the,Lazzeroni_Plaid_Models}.
\ignore{
, including iteratively clustering on subsets of the
genes/conditions~\cite{Getz_Coupled_two-way}, searching for
patterns/submatrices
of the expression data with low mean squared
residue[MEANING?]~\cite{Califano_Analysis_of, Cheng_Biclustering_of},
complementing sets of query genes
based on ranks[MEANING?]~\cite{Owen_Gene_Recommender}, fuzzy
clustering~\cite{Gasch_Exploring_the} and plaid
models[MEANING?]~\cite{Lazzeroni_Plaid_Models}.
[WE NEED TO SAY SOMETHING ABOUT WHY THESE APPROACHES ARE NOT AS GOOD AS
SA/ISA -
CAN WE SAY THAT THESE CLUSTERING METHODS DON'T DIRECTLY EXPLOIT THE
BIOLOGY, I.E.
REGULONS CONTROLLED BY TRANSCRIPTION FACTORS, WITH MANY TO MANY
MAPPING?]
}
A particularly promising approach, the Signature Algorithm (SA) was
introduced in 2002 by Ihmels {\it et al.}~\cite{SA}. Based on input sets
of related genes, SA identifies ``transcription modules'' (TMs), {\it
i.e.} sets of coregulated genes along with the sets of conditions for
which the genes are strongly coregulated. 
SA is well grounded in the biology of gene regulation. Typically, a
single transcription factor regulates multiple genes; a TM naturally
corresponds to a set of such genes and the conditions under which the
transcription factor is active. The authors tested the algorithm on a
large data set for the yeast {\it Saccharomyces cerevisiae}. By applying
SA to various sets of genes that were known or believed to be related,
they identified a large number of TMs.

Soon after, Bergmann {\it et al.}~\cite{ISA} published the Iterative
Signature Algorithm (ISA), which uses the output of SA as the input for
additional runs of SA until a fixed point is reached. By applying ISA to
random input sets and varying the threshold coefficient $t_{\rm{G}}$
(see below), the authors found almost all the TMs that had been
identified using SA, as well as a number of new modules. Many of these
modules proved to be in excellent agreement with existing knowledge of
yeast gene regulation.

While ISA can identify many transcriptional regulons from
gene-expression data, the algorithm has significant limitations. The
modules found depend strongly on the value of a threshold coefficient
$\tG$ used in the algorithm. To find all the relevant modules, a large
range of threshold values must be considered, and for each threshold the
algorithm may find thousands of fixed points, many of which are
spurious. While the largest, strongest modules are easily identified,
among the smaller, weaker modules it is a major challenge to identify
the real transcriptional regulons. Weak modules can even be completely
``absorbed" by stronger modules.

\begin{figure}
\includegraphics[width=3.375in]{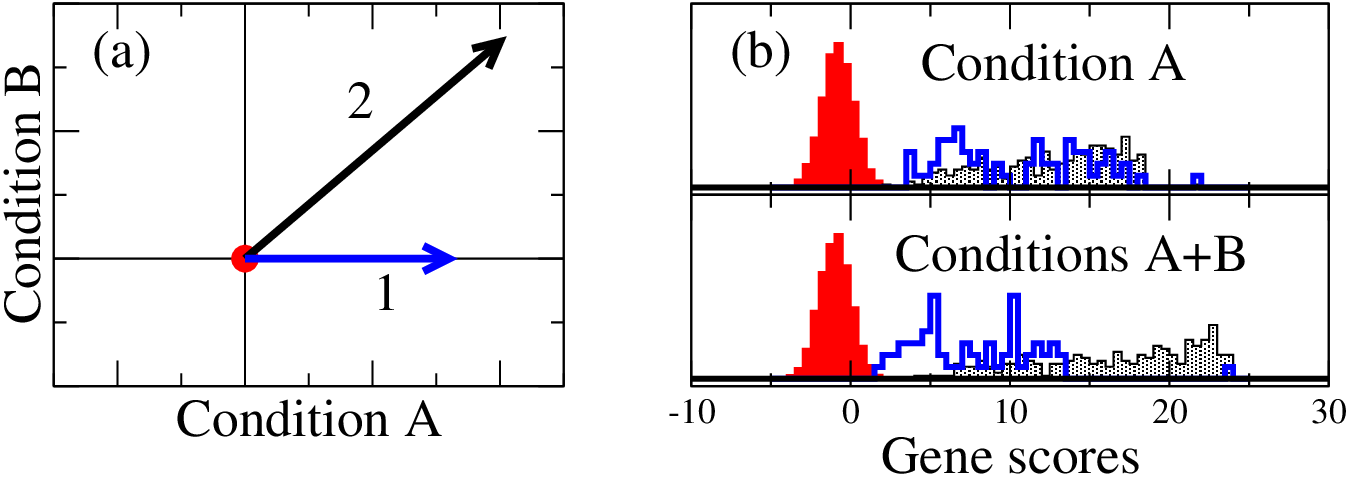}
\caption{
A toy model with only two transcription modules.  (a) Module 1 is
upregulated under condition A, while module 2, a larger, stronger
module, is upregulated under conditions A and B. (b) Normalized
histograms of the gene scores given by the Signature Algorithm (SA) for
the background (solid fill), module 1 (solid line) and module 2 (dotted
fill), when using the true condition vector for either module 1 (condition A)
or module 2 (conditions A+B). Even starting with the true condition
vectors, SA does not resolve the two modules. Nor
can the Iterated Signature Algorithm (ISA) resolve module 1, even if it
receives the module itself as input gene set, as the genes from module 2
have higher scores also for condition A (there is only one fixed point
of ISA). Due to the background noise, 
it is also impossible to separate the modules
by varying the ISA gene threshold coefficient $\tG$.}
\label{AbsorbedFixedPointFig}
\end{figure}

The performance limitations of ISA are related to a number of
algorithmic limitations. The need for a large range of thresholds is
partially due to the threshold definition, and the large number of fixed
points is due to the large positive feedback in the algorithm. The main
conceptual limitation of ISA, however, is that it only considers one
transcription module at a time. The algorithm does not use knowledge of
already identified modules to help it find new modules, and it may find
a strong module hundreds of times before it finds a given weak module.
An even worse case is shown in Fig.~\ref{AbsorbedFixedPointFig}: When a
strong and a weak module are coexpressed for a significant fraction of
conditions, it may be impossible to find the weak module by itself---ISA
will find only a single stable fixed point, dominated by the strong module.  

\begin{figure}
\includegraphics[width=3.375in]{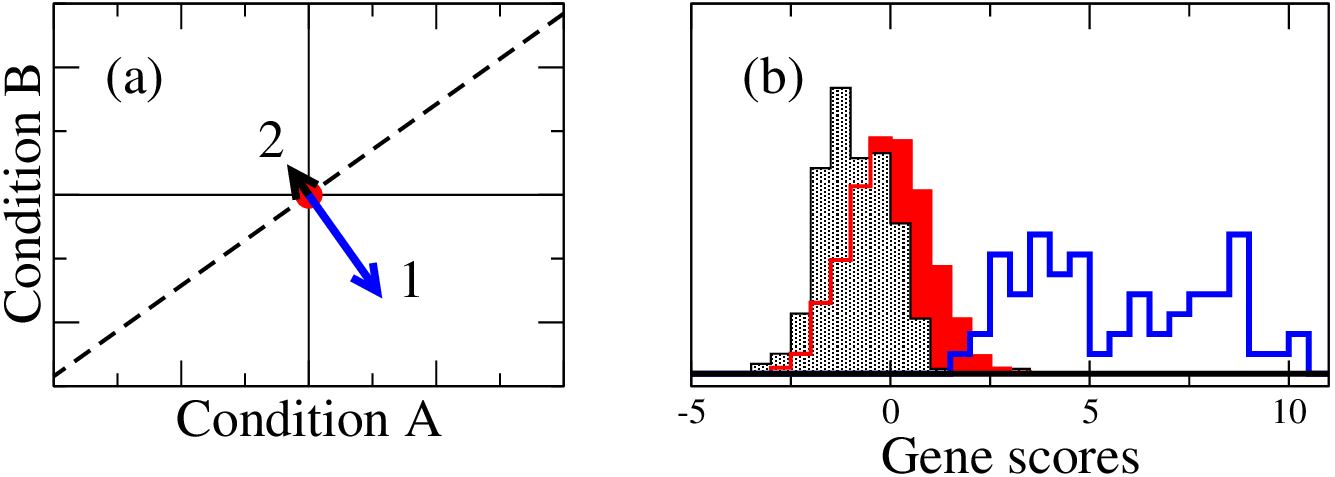}
\caption{Once the Progressive Itererative Signature Algorithm (PISA) has
eliminated the combined module 1+2 from Fig.~\ref{AbsorbedFixedPointFig}
(dashed line), the remaining signal makes it easy to separate the genes
of module 1 from the genes of module 2. (a) Remaining signal for each
module. (b) Actual gene scores for the new fixed point found by PISA.
Genes of module 1 (solid line) have been separated from genes of module
2 (dotted fill) and the background (solid fill).}
\label{PISAtoyex}
\end{figure}

A simple way to ensure that the same module is not found repeatedly is
to directly subtract the module from the expression
data, (this approach is used in~\cite{Lazzeroni_Plaid_Models}). A more
robust approach is to require the condition vector, {\it i.e.} the
weighted condition set, of each new transcription module to be
orthogonal to the condition vectors of all previously found modules. In
essence, this procedure corresponds to successively removing
transcription factors to reveal smaller and weaker transcription
modules. The successive removal of condition vectors is the central new
feature in our approach, and it is illustrated schematically in
Fig.~\ref{PISAtoyex}. We call the modified algorithm the Progressive
Iterative Signature Algorithm (PISA). Returning to the example in
Figs.~\ref{AbsorbedFixedPointFig}~and~\ref{PISAtoyex}, one finds that
PISA can easily identify both TMs: it first finds the strong module, 
removes its condition vector, and
then the only signal left is that of the weak module.

Progressively eliminating transcription modules {\it \`{a} la} PISA can
also improve the prospects for finding unrelated modules. The gene
regulation from one module will contribute to the
background noise for all unrelated modules. Therefore, eliminating large,
strong modules can significantly improve the signal to noise ratio of
the remaining modules.

\section {Methods}

\subsection {The Algorithms SA/ISA}

We briefly review the algorithms SA and ISA.
\ignore{The algorithm is based on the assumption that the logarithmic
gene expression ratios for the organism in question are, to a rough
approximation, given by the sum of contributions from a number of
transcription modules (plus noise):}
A transcription module $\Tm$ can be specified by a condition vector
(``experiment signature") $\tm^{\rm C}$ and a gene vector (``gene
signature") $\tm^{\rm G}$, where nonzero entries in the vectors indicate
conditions/genes that belong to the transcription module (TM).
\ignore{The (presumed) contribution of $\Tm$ to the expression level of
a gene $g$ for condition $c$ is then simply the product of the
corresponding vector entries: $({\bf \Delta E}(\Tm))_{gc}=(\tm^{\rm
C})_c (\tm^{\rm G})_g$.}

Given an appropriately normalized\footnote{SA actually uses two matrices
with different normalizations~\cite{SA}.} matrix $\mat{E}$ of log-ratio
gene expression data \comment{[MORTEN - IS "LOG-RATIO GENE-EXPRESSION
DATA" STANDARD TERMINOLOGY?]}
and an input set $G_{\rm I}$ of genes, SA scores all the conditions in
the data set according to how much each condition upregulates the genes
in the input set (downregulation gives a negative score). The result is
a condition-score vector $\vect{s}^{\rm C}$:
\beq
  \vect{s}^{\rm C} \equiv \frac{\mat{E}\transp \vect{m}_{\rm in}^{\rm
G}}{\abs{\vect{m}_{\rm in}^{\rm G}}},
\eeq
where $\mat{E}\transp$ is the transpose of $\mat{E}$ and
\beq
(\vect{m}_{\rm in}^{\rm{G}})_g=
  \left\{\begin{array}{ll} 1\;\;\;\;\;& g\in G_I \\
    0 & g\notin G_I\end{array}
  \right.
\eeq
is the gene vector corresponding to the input set. The entries of
$\vect{s}^{\rm C}$ that are above/below a threshold $\pm t_{\rm{C}}$
constitute the condition vector $\tm^{\rm C}$:
\beq
  (\vect{m}^{\rm C})_c \equiv (\vect{s}^{\rm C})_c
\cdot\Theta(\abs{(\vect{s}^{\rm C})_c}-t_{\rm C}),
\eeq
where $\Theta(x)=1$ for $x\geq 0$ and $\Theta(x)=0$ for $x<0$.

Similarly, the gene-score vector $\vect{s}^{\rm G}$ measures how much
each gene is upregulated by the conditions in $\tm^{\rm C}$, using the
entries of $\tm^{\rm C}$ as weights:
\beq
  \vect{s}^{\rm G} \equiv \frac{\mat{E}\; \vect{m}^{\rm
C}}{\abs{\vect{m}^{\rm C}}}.
  \label{GeneScoreEq}
\eeq
The entries of the gene-score vector $\vect{s}^{\rm G}$ that are more
than $\tG$ standard deviations $\sigma_{\vect{s}^{\rm G}}^{_{}}$ above
the mean gene score in the vector $\vect{s}^{\rm G}$ constitute the gene
vector $\tm^{\rm G}$:
\beq
  (\vect{m}^{\rm G})_g \equiv (\vect{s}^{\rm G})_g
\cdot\Theta((\vect{s}^{\rm G})_g-\avg{(\vect{s}^{\rm G})_g}_g-t_{\rm
G}\sigma_{\vect{s}^{\rm G}}^{_{}})
\eeq

ISA uses $\vect{m}^{\rm G}$ as the input $\vect{m}_{\rm in}^{\rm G}$ for
the next iteration, {\it i.e.} the genes are now weighted according to
their gene scores, until a fixed point is reached.

\subsection {The Algorithm PISA}

\noindent {\it Orthogonalization.}
Within PISA, each condition-score vector $\vect{s}^{\rm C}$ is required
to be orthogonal to the condition-score vectors of all previously found
transcription modules (TMs). Therefore, whenever PISA finds a TM and its
associated condition-score vector $\vect{s}^{\rm C}$, the component
along $\vect{s}^{\rm C}$ of each gene is removed from the gene
expression matrix (see {\it Implementation of PISA} below). This
requirement of orthogonality in PISA conflicts with the condition-score
threshold as used in ISA. If we make the condition-score vector
orthogonal first and then apply the threshold, the vector will no longer
be orthogonal, whereas if we apply the threshold first,
orthogonalization will give nonzero weight to all conditions,
eliminating the noise-filtering benefit of thresholding. We have chosen
to eliminate the condition-score threshold completely. In any event,
conditions that in ISA would fall below the threshold will have low
weight and will give only a small contribution to the noise.\\

\begin{figure}
\includegraphics[width=3.375in]{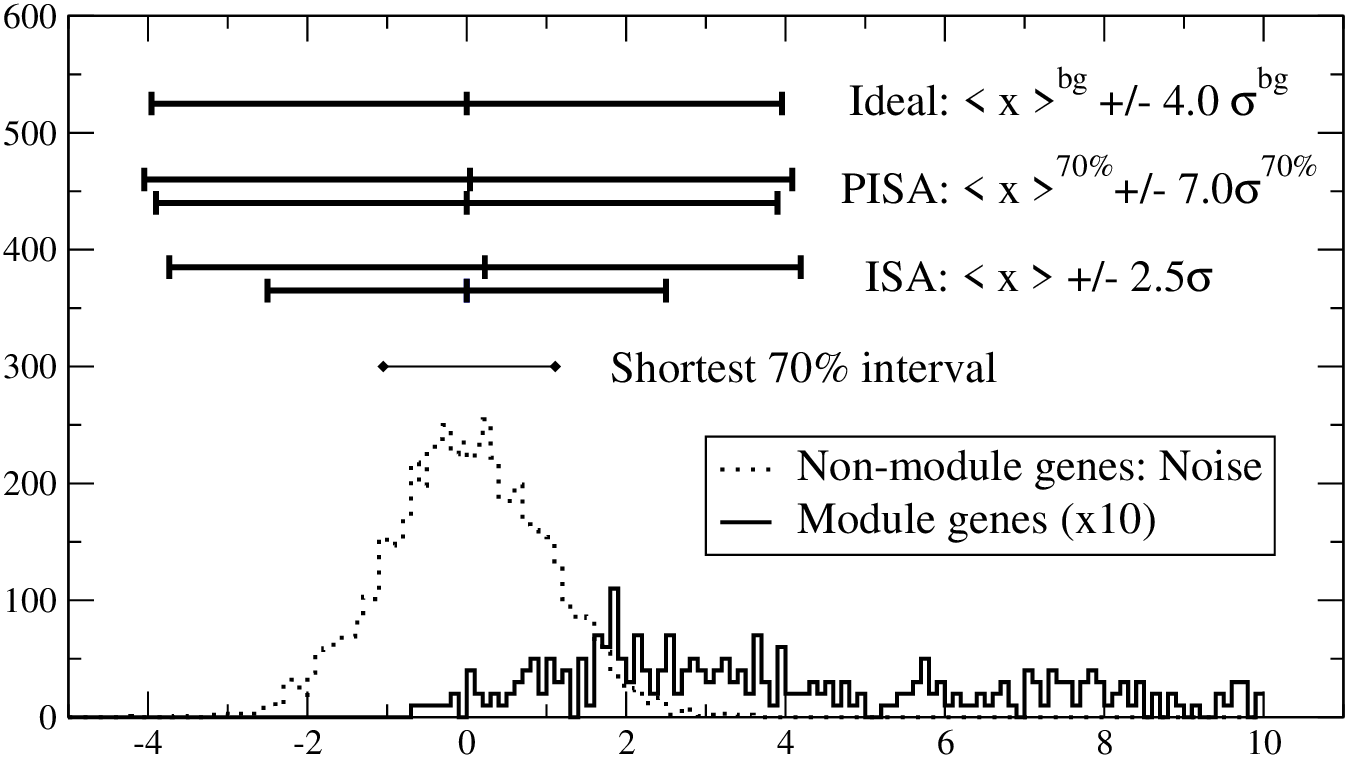}
\caption{Means and standard deviations as used in ISA and in PISA
algorithms, calculated using all the genes (top bars) or only the non-module
genes (bottom bars). The mean $\avg{x}^{70\%}$ and standard deviation
$\sigma^{70\%}$ from PISA, using only the distribution within the
shortest interval that contains 70\% of all genes, are almost identical
to the ideal values $\avg{x}^{\rm bg}$ and $0.56\sigma^{\rm bg}$ of the
background noise (non-module genes). In ISA, the mean and standard
deviation are calculated from the whole distribution and so are strongly
module dependent.
\ignore{ is much larger. Within ISA, to get an effective threshold of
4.0 $\sigma^{\rm bg}$ the nominal threshold must be $\approx 2.5
\sigma^{\rm ISA}$.}
This example uses generated data for a module of 300 genes out of 6206
total genes. The non-module genes have a normal distribution of
gene-expression levels.}
\label{geneeffectivethresholdfig}
\end{figure}

\noindent {\it The gene-score threshold.}
In ISA, to find all modules, it is necessary to run the algorithm with
many different threshold coefficients $\tG$.
\ignore{(in~\cite{ISA}, the authors used thresholds from 1.8 to 4.0,
but this is not enough to find all possible modules).}
For low thresholds one finds a few very large modules (many genes),
while for high thresholds one finds many small modules (few genes).
Without prior knowledge of the module one is searching for, it is
difficult to know what $\tG$ to use.
\ignore{This raises the issue of which value of the threshold to
use for PISA---}
Within PISA, we wish to find all the modules using a single threshold.
\ignore{which appears to be difficult. Fortunately, this problem is to a
large extent caused by}
This requires modifying the threshold definition. In ISA, the gene-score
threshold is $t_G \sigma^{\rm ISA}$ where the standard deviation
$\sigma^{\rm ISA}$ is computed using the full distribution of gene
scores, and includes contributions both from the background and from the
module of interest (Fig.~\ref{geneeffectivethresholdfig}). For large,
strong modules, the module contribution may be larger than the
background contribution. As a result, $\sigma^{\rm ISA}$ is module
dependent and $t_G$ must be adjusted to prevent false-positives from the
background.

We eliminate this problem in PISA by specifying the threshold relative
to the background, which we estimate using the mean, $\avg{x}^{\rm
70\%}$, and the standard deviation, $\sigma^{\rm 70\%}$, of the gene
scores within the shortest interval that contains at least 70\% of all
the gene scores. By excluding extreme gene scores in this way, we
minimize the influence of the module on the means and standard
deviations of gene scores (Fig.~\ref{geneeffectivethresholdfig}). As a
test, we used $\sigma^{\rm 70\%}$ in place of $\sigma^{\rm ISA}$ in ISA
and found both very large and very small modules with a single value of
$t_G$.

We need to be conservative when selecting the gene-score threshold
because, if PISA misidentifies a module, elimination of its condition
vector can lead to errors in other modules. Therefore, the number of
genes included in modules due to noise should be very low. We have used
a threshold of $7.0\sigma^{\rm 70\%}$, which for a Gaussian distribution
corresponds to about $3.9\sigma$. The chance of including a gene due to
noise is about $10^{-4}$ per gene, {\it e.g.} with the 6206 genes in the
yeast data set, the average number of genes included by mistake in each
module would be about 0.62. Using a high threshold means that we may
miss genes that should belong to a module, however this is less risky
than including genes by mistake. As PISA proceeds by eliminating
condition-score vectors, it does not matter whether we identify all the
genes in a module, as long as the condition-score vector is accurate.
Once,
PISA has finished, we can easily see which genes would be included
when using various gene-score thresholds for the same condition-score
vector.

ISA only considers sets of genes that have {\it high} gene scores, {\it
i.e.} positive signs. As discussed in~\cite{SA}, this can lead to two
modules that are regulated by the same conditions but with opposite
sign. In contrast, PISA includes all genes with sufficiently extreme
scores in a single module, and the relative signs of gene scores specify
whether the genes are coregulated or counter-regulated.\\

\noindent {\it Implementation of PISA.}
To begin, PISA requires a matrix {\bf E} of log-ratio gene-expression
data, with zero average for each condition.
Two matrices are obtained from {\bf E}: The first $\Eg$ is
normalized for each gene 
$$\avg{(\Eg)_{gc}}_c=0,\;\;
\avg{(\Eg)_{gc}^2}_c=1\;\;\;\;\;\;\forall g\in G.$$
Normalization of $\Eg$ is essential so that the gene-score threshold
can be applied to all genes on an equal footing. 
The second matrix $\Eort$ is obtained from $\Eg$ by normalizing for
each condition, $ \avg{(\Eorti)_{gc}^2}_g=1$, 
where $\Eorti$ denotes the initial $\Eort$.
(Note that this is essentially the opposite of the notation used
in~\cite{SA}.)
We then apply a modified version of ISA, mISA (see below), a large
number of times (typically 10,000), and whenever mISA finds a module, we
remove from $\Eort$ the components along the module's condition score vector $\vect{s}^{\rm C}$:
\beq
  \Eort^{\mr{new}} \equiv \Eort- \Eort \frac{\vect{s}^{\rm{C}}
(\vect{s}^{\rm{C}})\transp}{\abs{\vect{s}^{\rm{C}}}^2}
\eeq
\ignore{The new matrix $\Eort^{\rm new}$, which is used for the next
applications of mISA, will still have zero average over genes $g$,
conditions $c$.}

\ignore {The matrix $\Eg$, which is used to calculate the gene scores,
stays properly normalized throughout the algorithm, thus the scores for
different genes will always be comparable. As we don't use a condition
threshold, we don't have to worry about whether or not the condition
scores, calculated using $\Eort$, are comparable or not.}

As mISA is repeatedly applied, new modules are found less and less
frequently. For example, one run of 10,000 applications of mISA found
496 modules, and 287 of them were found in the first 1,000 applications.
\ignore{711 modules, and 504 of them were found in the first 2,000} 
As the later modules are also generally smaller and less reliable, the
exact number of times mISA is applied is not very important.\\

\noindent {\it mISA.}
As input, the modified Iterative Signature Algorithm (mISA) requires the
two matrices $\Eort$ and $\Eg$. We start each application of mISA by
generating a random set of genes $G_0$ and a corresponding gene vector
$\vect{m}_0^{\rm{G}}$:
\ignore{The number of genes in $G_0$ is currently a random number $2\leq
\abs{G_0}\leq 51$.}

$$(\vect{m}_0^{\rm{G}})_g=\left\{\begin{array}{ll} 1\;\;\;\;\;& g\in G_0
\\ 0 & 
g\notin 
G_0\end{array} \right.$$
Each iteration $i$ within mISA consists of multiplying the transpose of
$\Eort$ by the gene vector $\vect{m}_i^{\rm G}$ to produce the
condition-score vector $\vect{s}_i^{\rm C}$:
$$\vect{s}_i^{\rm{C}} \equiv \Eort\transp \vect{m}_{i}^{\rm{G}},$$
and then multiplying $\mat{E}_0$ by the normalized condition-score
vector to produce the gene-score vector $\vect{s}_i^{\rm G}$:
$$\vect{s}_i^{\rm{G}} \equiv \frac{\Eg 
\vect{s}_i^{\rm{C}}}{\abs{\vect{s}_i^{\rm{C}}}},$$
From $\vect{s}_i^{\rm G}$, one calculates the gene vector
$\vect{m}_{i+1}^{\rm G}$ for the next iteration:
$$(\vect{m}_{i+1}^{\rm{G}})_g \equiv (\vect{s}_i^{\rm{G}})_g\, 
\theta(|(\vect{s}_i^{\rm{G}})_g-\avg{(\vect{s}_i^{\rm{G}})_g}_g^{70\%}|-t_{\rm{G
}} 
\sigma_{\vect{s}^{\rm{G}}_i}^{70\%})$$

We iterate until: (a) $(\vect{m}^{\rm{G}}_i)_g$ and $(\vect{m}^{\rm{G}}_{i+1})_g$ have the same sign (0, + or -) for all $g$, (b) the iteration number is $i=20$, or (c) fewer than two genes have nonzero weight. If fewer than five genes have nonzero weight (for (a) or (b)), the result is discarded, otherwise we have found a module with condition-score vector
$\vect{s}^{\rm{C}}=\vect{s}^{\rm{C}}_i$, gene-score vector
$\vect{s}^{\rm{G}}=\vect{s}^{\rm{G}}_i$,  and gene vector
$\tm^{\rm{G}}=\vect{m}^{\rm{G}}_{i+1}$.

We chose a threshold coefficient $t_{\rm{G}} = 7.0$ so that the expected
number of genes included in each module due to background noise would be
less than one. However, with this high threshold, starting from a random
set of genes there was only a very low chance that two or more genes
would score above the threshold in the first iteration\footnote{This is
not an issue in ISA, where the condition threshold helps to pick out
the, possibly very small, signal from the noise.}. To increase the
chance of finding a module, we used a different formula for
$\vect{m}_{1}^{\rm{G}}$. Instead of selecting only genes with scores
above the threshold, we kept a random number $2\leq n\leq 51$ of the
genes with the most extreme scores. This procedure was generally
adequate to produce a correlated set of genes for the next iteration.\\

\noindent {\it Consistent modules.}
ISA typically finds many different fixed points corresponding to the
same module, each differing by a few genes. PISA only finds each module
once during a run, but the precise genes in the module depend on the
random input set of genes and also on which modules were already found
and eliminated. Furthermore, PISA sometimes finds a module by itself,
while other times it may find the module joined with another module, or
PISA may find only part of a module, or not find the module at all. To
get a reliable set of modules, it was necessary to perform a number of
runs of PISA and identify the modules that were consistent from run to
run. To identify consistent modules, we first tabulated preliminary
modules -- transcription modules found by individual runs of PISA. A
preliminary module contributes to a consistent module if the preliminary
module 
contains more than half the genes in the full module, regardless of
gene-score sign, 
and these genes constitute at least 20\% of the genes in the preliminary
module.
A gene is included in the consistent module if the gene occurs in more
than 50\% of the contributing
preliminary modules, always with the same gene-score sign. \\

\noindent{\it Correlations between condition-score vectors.}
Once we identified a consistent module, $\vect{m}^{\rm G}$, we
calculated the raw condition-score vector $\vect{r}=\Eorti \transp
\vect{m}^{\rm G}$, using the initial value of the gene-expression-data
matrix $\Eort$. From the $\vect{r}$'s we evaluated the condition
correlations
$\vect{r}\cdot\vect{r}^\prime/(\abs{\vect{r}}\abs{\vect{r}^\prime})$
between different modules. 
\ignore{(This correlation can 
between individual genes from the two modules, as we here average over
many genes, reducing the noise.)}
\ignore{(This correlation can be much larger that the average
correlation between individual genes from the two modules: For the
modules we have found for (1) iron (and other metals) metabolism and (2)
phosphate metabolism, the average correlation between genes is 0.0135,
while the correlation between the modules is 0.11. These values,
however, depend on the normalization used.)}
\\

\noindent Additional details are discussed in the supporting material.
\ignore{\noindent {\it Details.}[I'D PREFER TO SHORTEN THIS TO JUST
"ADDITIONAL
DETAILS ARE DISCUSSED IN 
THE SUPPORTING MATERIAL". INSTEAD OF PUTTING THIS IN ITS OWN SECTION,
CAN IT GO AT THE BOTTOM OF
THE IMPLEMENTATION OF PISA SUBSECTION?]
There are some additional, relatively minor changes that we have made to
ISA,either to improve performance or to compensate for removing the
threshold $t_{\rm C}$. These are discussed in the supporting material.}

\subsection{$p$-Values}

Given a set containing $m$ genes out of the total of $N_{\rm G}$, the
$p$-value for having at least $n$ genes in common with a Gene Ontology
(GO) category containing $c$ of the $N_{\rm G}$ genes is
\beq
  p = \sum_{i=n}^{\min\{c,m\}} \frac{{{c}\choose{i}}{{N_{\rm
G}-c}\choose{m-i}}}{{{N_{\rm G}}\choose{m}}},
\eeq
\ignore{\beq
  p = 1-\sum_{i=0}^{n-1} \frac{{{c}\choose{i}}{{N_{\rm
G}-c}\choose{m-i}}}{{{N_{\rm G}}\choose{m}}}.
\eeq}
We ignore any genes that are not present in our expression data when counting $c$.

\section{Results}

We applied PISA to the yeast data set used in~\cite{ISA}, which consists
of log-ratio gene-expression data for $N_{\rm G}=6206$ genes and $N_{\rm
C}=1011$ experimental conditions (approximately 10\% of the values are
missing or invalid). Normalization gives the matrices $\Eg$ and $\Eort$
(see Methods for details).

As a preliminary test, we repeatedly applied PISA to one fully scrambled
version of the matrix $\Eg$ 
(and the corresponding $\Eort$). From run to run, the algorithm
identified many large modules derived almost entirely 
from a single condition, as expected in light of the broad distribution
of the raw gene-expression data (Fig.~\ref{distributions} in supporting material). 
\ignore{This is not too surprising, as the distributions for raw expression
data
that derive from only a very small number of 
statistics will be far from Gaussian (for most modules, the gene scores
are averages of many 
}
PISA also found many small modules, but these differed from one run to
the next.
We were able to eliminate both these classes of false positives using
filters for consistency, recurrence, and number of contributing
conditions (Fig.~\ref{modulequality} in supporting material).

We performed 30 runs of PISA on the yeast data set
\ignore{(the same data set used in~\cite{ISA})}
and identified the modules that appeared consistently, using the filters
derived above.
At the start of each run, only a few modules could be found with our
single choice of gene threshold $\tG$. Nevertheless, PISA did
consistently find new modules after eliminating others, demonstrating
that removing the condition vectors of found modules improves the 
signal to noise for the remaining ones.

For most of the modules we found, the genes were coregulated, {\it
i.e.} all the gene scores had the same sign. (In contrast, the modules
that were eliminated by the filters often had about equal numbers of
genes of either sign.) There were, however, a significant number of
modules with a few gene scores differing in sign from the rest, and a
few modules 
with many gene scores of both signs, {\it e.g.} the a/$\alpha$ pheromone
production/detection module. Furthermore, many of the modules found by
PISA agreed closely with modules identified by ISA at various
thresholds, while other PISA modules were subsets of ISA modules. Some
PISA modules, for example the de novo purine synthesis module
(Fig.~\ref{PurineModule}), were significantly more complete than the
ones found by ISA (at any threshold).

\begin{figure}
\module{
\moduletitlesize

Module: De novo purine biosynthesis
\modulesize

\vspace{10pt}
Number of genes: 32

Average number of contributing conditions: 14.6

Consistency: 0.83

Best ISA overlap: 0.59 at threshold 5.0, frequency 16

\vspace{10pt}
\genesize
\begin{picture}(300,150)(0,0)
\ignore{
       MTD1  28  490.002
       ADE2  28  478.121

       SHM2  28  468.812
      ADE17  28  446.213
      ADE13  28  419.476
       ADE1  28  414.622
       GCV1  28  403.635
       ADE4  28  339.101
     ADE5,7  26  334.772
       CEM1  28  314.833
       GCV2  28  310.791
       SER1  28  310.702
    YGL186C  28  310.375
       GCV3  28  286.458
      ADE12  28  284.974
       HIS4  28  273.873
       ADE8  27  265.844
    YDR089W  27  243.694
       ADE6  25  235.064
       SER2  26  231.39
       SER3  26  228.487
      SER33  27  222.383
       ADE3  26  217.567
       MET6  26  204.301
    YPR004C  25  203.972
       AIP2  24  186.574
   ETF-BETA  23  153.927
       HIS7  21  135.631
       URA4  21  128.485
       SSU1  18  115.747
       HIS1  17  108.011
       HIS5  16  96.4565
}
\put(  65,  75){\gene{2}{GCV3}}
\put( 325, 125){\gene{1}{ADE1}}
\put( 195,  25){\gene{3}{HIS7}}
\put( 195,  75){\gene{3}{HIS4}}
\put(  65,  25){\gene{4}{AIP2}}
\put(   0, 100){\gene{2}{GCV1}}
\put( 325,  75){\gene{0}{YDR089W}}
\put( 260,  75){\gene{1}{ADE8}}
\put(   0,   0){\gene{3}{HIS1}}
\put( 195, 100){\gene{4}{CEM1}}
\put( 130,  50){\gene{2}{SER3}}
\put( 325,  50){\gene{2}{MET6}}
\put(   0,  75){\gene{1}{YGL186C}}
\put( 130, 100){\gene{1}{ADE5,7}}
\put(   0,  50){\gene{1}{ADE6}}
\put( 260,  50){\gene{1}{ADE3}}
\put( 130,  25){\gene{4}{ETF-BETA}}
\put(  65,  50){\gene{2}{SER2}}
\put( 195,  50){\gene{2}{SER33}}
\put(  65,   0){\gene{3}{HIS5}}
\put(   0, 125){\gene{2}{MTD1}}
\put( 130, 125){\gene{2}{SHM2}}
\put( 260, 125){\gene{1}{ADE13}}
\put( 260,  25){\gene{4}{URA4}}
\put( 195, 125){\gene{1}{ADE17}}
\put( 260, 100){\gene{2}{GCV2}}
\put(  65, 100){\gene{1}{ADE4}}
\put( 130,  75){\gene{1}{ADE12}}
\put(  65, 125){\gene{1}{ADE2}}
\put( 325, 100){\gene{2}{SER1}}
\put( 325,  25){\gene{4}{SSU1}}
\put(   0,  25){\gene{4}{YPR004C}}
\end{picture}

\vspace{20pt}
\gene{0}{0} \modulesize Unknown \genesize

\vspace{5pt}
\gene{1}{1} \modulesize Purine synthesis/transport \genesize

\vspace{5pt}
\gene{2}{2} \modulesize Tetrahydrofolate activation \genesize

\vspace{5pt}
\gene{3}{3} \modulesize Histidine biosynthesis \genesize

\vspace{5pt}
\gene{4}{4} \modulesize Other \genesize

\vspace{10pt}
\includegraphics[width=3.2in]{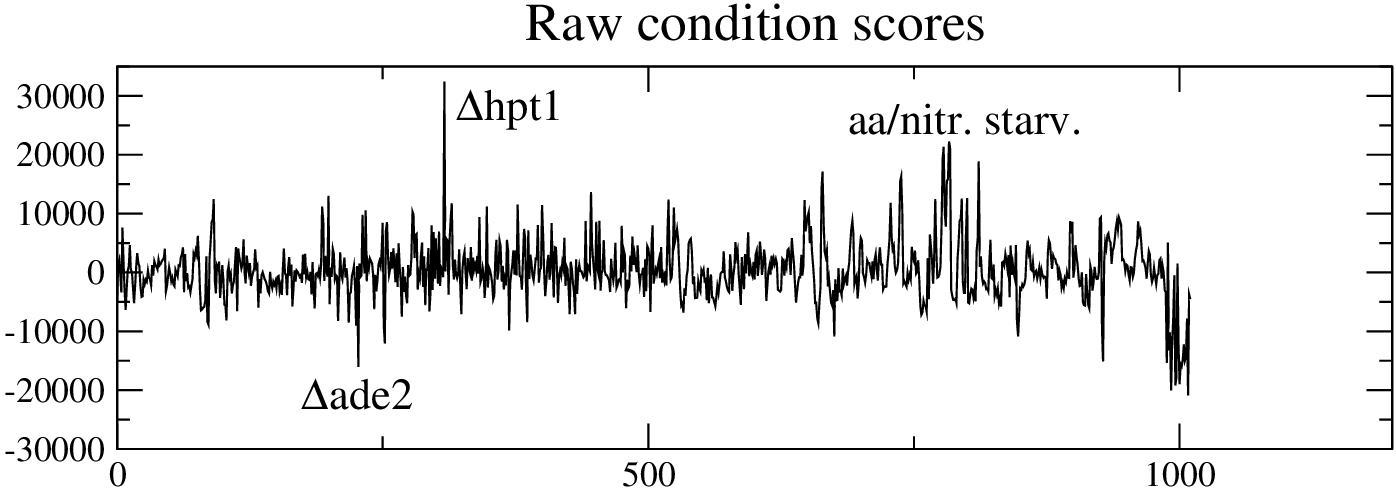}
}
\caption{The de novo purine synthesis module found with PISA.}
\label{PurineModule}
\end{figure}

\ignore{
We also find many modules that derive almost entirely from a single
condition. For these modules, the statistical propertiThis means that,
once the earlier modules had eliminated, the expression pattern caused
by this mutation was not similar to any other conditions. These modules
are not very reliable (as the the gene scores are hardly averaged at
all), and certainly don't have much relevance under physiological
conditions.
}

\begin{figure}
\begin{picture}(240,160)(0,0)
\put(0,0){\includegraphics[width=3.375in]{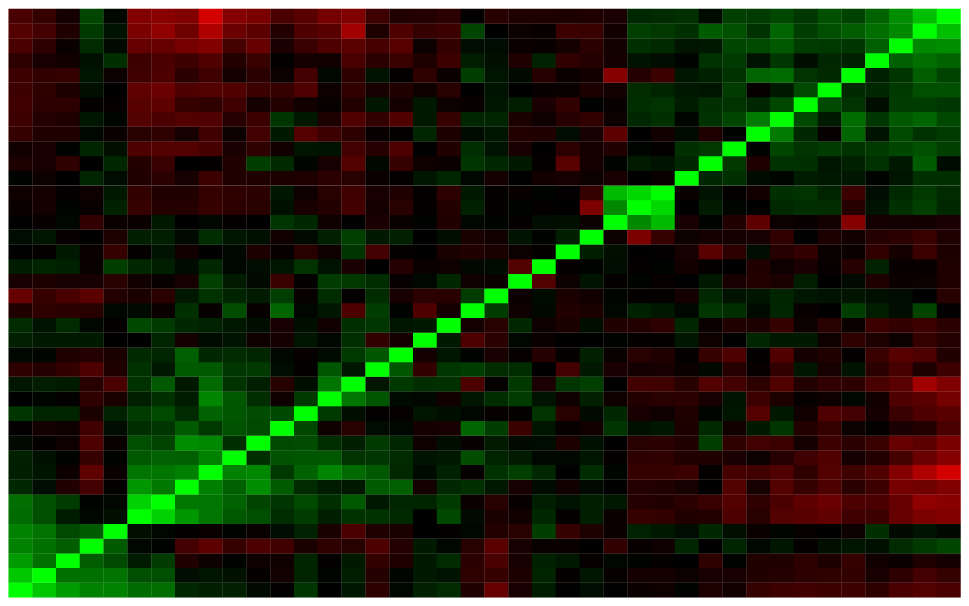}}
\comment{\multiput (0,0)(0,10){16}{\line(1,0){2}}
\multiput (0,0)(10,0){24}{\line(0,1){2}}}
\put(42,10){\moduleannotation{$\left.\mbox{\vspac{10}}\right\}$Biosynthesis}}
\put(105,3.5){\moduleannotation{-Amino acid (general)}} 
\put(190,7.1){\moduleannotation{-Arginine}}
\put(105,10.7){\moduleannotation{-Biotin}}
\put(190,14.3){\moduleannotation{-Lysine}}
\put(105,17.9){\moduleannotation{-De novo purine}}
\put(90,31){\moduleannotation{$\left.\mbox{\vspac{14}}\right\}$}}
\put(98,22){\rotatebox{90}{\moduleannotation{Stress}}}
\put(170,21.5){\moduleannotation{-Oxidative stress}}
\put(170,28.6){\moduleannotation{-Proteolysis}}
\put(170,35.8){\moduleannotation{-COS genes}}
\put(170,43.0){\moduleannotation{-S-S bond repair}}
\put(110,25.1){\moduleannotation{-AAD genes}}
\put(110,32.2){\moduleannotation{-Trehalose++}}
\put(110,39.4){\moduleannotation{-Heat shock}}
\put(105,95){\moduleannotation{Mating$\left\{\mbox{\vspac{8}}\right.$}}
\put(10,89.7){\moduleannotation{\makebox[90pt][r]{$\alpha$/a
difference-}}}
\put(10,96.8){\moduleannotation{\makebox[90pt][r]{Mating type a
genes-}}}
\end{picture}
\caption{Correlations between modules identified by PISA. The modules
are ordered to form clusters: In the lower left corner are the main
amino acid biosynthesis module and several smaller, more specific
biosynthesis modules; the other main clusters are, roughly, stress
response, mating, and ribosomal proteins/rRNA processing.}
\label{Correlations}
\end{figure}

\ignore{Examples of correlated modules are: ribosomal proteins (104
genes) and RNA-related genes (144 genes)---correlation 0.76 and with no
common genes (in ISA, these modules are combined at sufficiently low
$\tG$). The protein synthesis module (104 genes) is correlated both with
de novo purine synthesis (28 genes) with 6 common
genes and a correlation of 0.51, and with nitrogen/sulfur metabolism (58
genes, correlation0.57), but the correlation between the last two
modules is only 0.30, {\it i.e.} they seem to be related only through
the protein-synthesis module.
}

\ignore{
The galactose induced module turns on GAL genes and also, as a weaker
effect, represses a number of hexose transporters. These hexose
transporters also occur in another module (which is consistently found
after the galactose induced module) which consists almost entirely of
hexose transporters, and this module also contains GAL2, the galactose
permease, but in this module it is coregulated with the other hexose
transporters, whereas they were counter-regulated in the galactose
induced module.
}

PISA found several small modules that agree very well with 
known gene regulation in yeast. For example, the arginine-biosynthesis
module
consists of ARG1, ARG3, ARG5,6, ARG8, CPA1, CTF13, and CAR2; out of
these
CAR2 has a negative gene score, {\it i.e.} it is counter-regulated
relative
to the others. The first five genes are precisely the arginine-synthesis
genes known to be repressed by arginine, while CAR2 and CAR1
(which is the 2nd highest scoring gene that failed to make the
threshold)
are catabolic genes known to be induced by
arginine~\cite{Messenguy_Regulation_Of}.

PISA also found a
zinc (zap1 regulated) module consisting of ZRT1, ZRT2, ZRT3, ZAP1,
YOL154, INO1, ADH4, and YNL254C. These are almost exactly the highest
scoring genes in a microarray experiment comparing expression under
zinc starvation of a zap1 mutant vs. wild
type~\cite{Lyons_Genome-wide_characterization}, however, our data set
does not include this or any other zinc starvation (or zap1 mutant)
experiment---indeed, there are no experimental conditions that have a
remarkably high score for this module, although conditions from the
Rosetta compendium~\cite{Rosetta}, most of which are deletion mutant
experiments, tend to have much higher scores than the other conditions
(see supporting material). This module, as well as the starvation
experiments in~\cite{Lyons_Genome-wide_characterization} and direct
transcription factor binding experiments (see below), all indicate that
YNL254C is regulated by zap1, and it probably has some function related
to zinc starvation/uptake.

\begin{figure}
\includegraphics[width=3.375in]{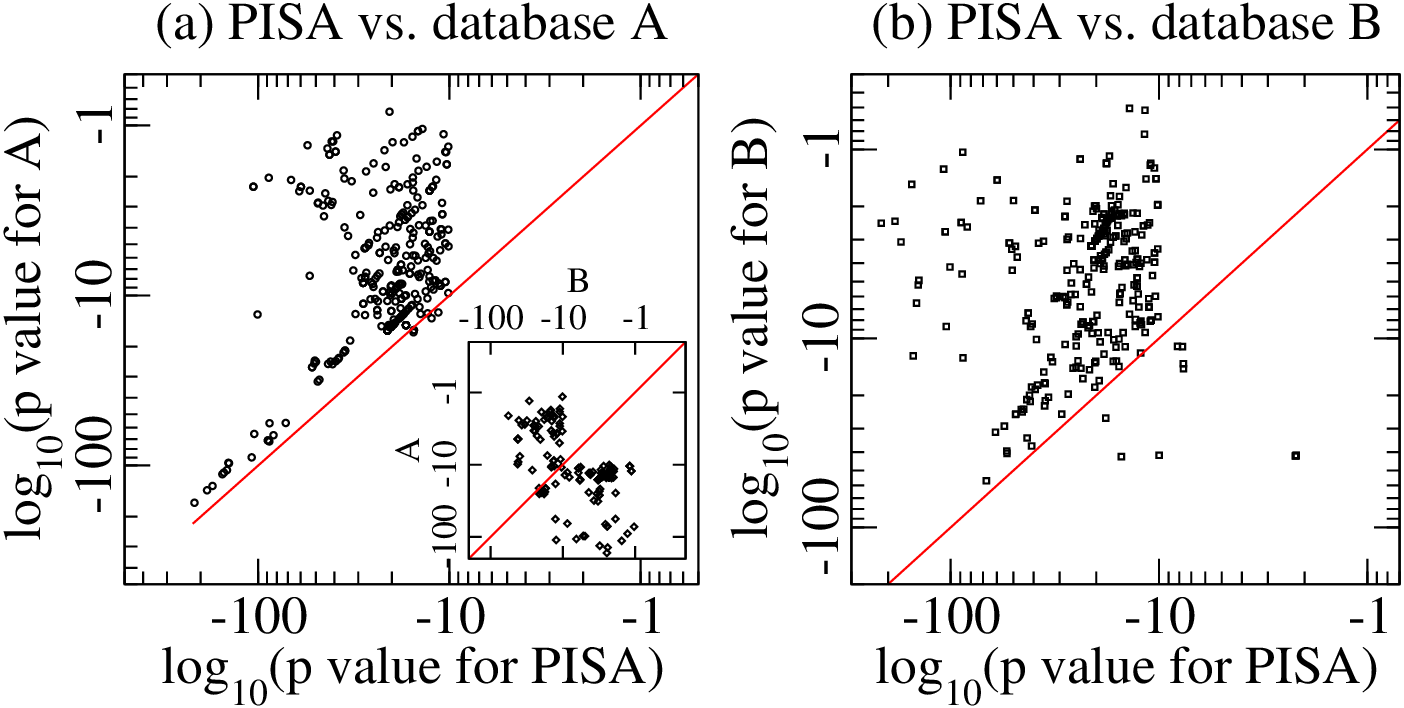}
\caption{Best $p$-values onto every Gene Ontology (GO) category.
In each panel, we include only GO categories for which at least one
$p$-value is below
$10^{-10}$. (a) PISA vs. Database A. (b) PISA vs. Database B. (a) inset:
Database A vs.
database B---there are very few GO categories onto which both A and B
have
low $p$-values.}
\label{GOcomparisons}
\end{figure}

In order to evaluate the overall performance of PISA, we compared our
modules to the categories in the Gene Ontology (GO) curated 
database~\cite{GO}. For the set of genes in each of our modules 
we calculated the $p$-value for the overlap with the set of genes in
every GO 
category (see Methods). The $p$-value is the probability that an 
observed overlap occurred by chance. The lowest $p$-value we found was
$3.5\cdot 10^{-216}$, for the GO category ``cytosolic ribosome'', 
and we found $p$-values
below $10^{-20}$ for more than 140 other GO categories. (The modules
that were
removed by our filters mostly did not have significant $p$-values, and
none were below $10^{-10}$).
We used the p-values between our PISA modules and the GO categories to
compare PISA 
to other means of identifying transcriptional modules. Specifically, we 
compared PISA to two different databases of genes predicted to be
regulated by single transcription factors. Database ``A'' contains
genes that were enriched through immunoprecipitation with tagged
transcriptional regulators~\cite{Lee_Transcriptional_Regulatory}, while
Database ``B'' has genes sharing regulatory sequences derived by
comparative genomics~\cite{Kellis_Sequencing_and}.
Figure~\ref{GOcomparisons} shows the $p$-values between GO and PISA
compared to
the $p$-values between GO and each of these two databases.\footnote{We
used an internal $p$-value threshold of
0.001 for Database A, as suggested
in~\cite{Lee_Transcriptional_Regulatory}.} The lower $p$-values for PISA
indicate a consistently
better agreement between GO and PISA than between GO and the other
databases. For a few GO categories
Database B has a lower $p$-value than PISA, but these categories are all
close to the root of the GO tree 
and each contains more than half the genes in yeast.

Compared to microarray data, both Database A and Database B 
have a clear disadvantage: their binding sites are assigned to
intergenic
regions, and if the two genes bordering an intergenic region are
divergently transcribed, then the databases do not identify which of the 
genes is regulated. In many cases, we found that by comparing sets of
genes in database A to PISA modules, we
could decide which of divergently transcribed genes were
actually regulated. For example, Database A lists 6 intergenic regions
as binding site for zap1 at an internal $p$-value threshold of $10^{-5}$,
and 4 of these lie 
between divergently transcribed genes. However, 5 of the 6 intergenic
regions border the genes
ZRT1, ZRT2, ZRT3, ZAP1, and YNL254C which PISA identifies as part of the
zinc module.

Database A appears to have an additional source of false positives.
Intergenic regions that are close to intergenic regions with very low
$p$-values often have low $p$-values themselves, even when there is no
apparent connection between the genes and no evidence of a binding site in
the DNA sequence.  For example, for the de novo purine-biosynthesis
module, which is primarily regulated by the bas1 transcription factor,
the intergenic region controlling GCV2 has the lowest $p$-value within
Database A, $1.1\cdot 10^{-16}$, and all the four closest intergenic 
regions have $p$-values below $10^{-5}$. Comparison to PISA 
modules can help eliminate these potential false positives: 
out of the 29 genes assigned a $p$-value below $10^{-4}$ for bas1 
binding in database A, 13 belong to a single PISA module, 
4 others are divergently transcribed adjacent genes, and 6 others 
are genes transcribed from nearby intergenic regions. 


\ignore{
The agreement between the PISA modules and the two databases was mixed,
however the PISA modules agreed significantly better with database B
than the two databases did with each other. (Note that while both of the
databases attempt to identify genes that bind single transcription
factors, the PISA modules are effective modules, i.e. they may involve
combinatorial patterns of many transcription factors, and we should not
expect perfect agreement with the databases.) While some modules show
excellent agreement with database B (e.g. the ribosomal proteins
module), other modules (e.g. phosphate starvation), while clearly very
good, have no significant overlap with the databases. The agreement with
the database B was much better for the modules that we kept than for the
modules that were discarded by the filters: The geometric average of the
probability of the most significant overlap was $10^{-6}$ for the
modules we kept, but only 1/58 for the modules that were discarded (when
comparing to 113 different database sets).
}

\section{Discussion}

The Progressive Iterative Signature Algorithm
(PISA) embodies a new approach to analysis of large gene-expression 
data sets. The central new feature in PISA is the robust elimination 
of transcription modules as they are found, by removing their 
condition-score vectors. Also new to PISA, compared to its 
precursors SA \cite{SA} and ISA \cite{ISA}, is the inclusion of both 
coregulated and counter-regulated genes in a single module, and the
use of a single gene-score threshold.

Altogether, these new features result in an algorithm that 
can reliably identify both large and small regulatory modules, 
without supervision. We confirmed the performance of PISA by
comparison to the Gene Ontology (GO) database -- PISA performed
considerably better against GO than either high-throughput 
binding experiments or comparative genomics. PISA therefore 
provides a practical means to identify new regulatory modules
and to add new genes to known modules. 


Can PISA shed any light on the organization of gene expression 
beyond the level of individual transcription modules?
In~\cite{ISA}, the authors argued that they could trace the relationship
between modules from the effects of changing the threshold $\tG$. For
instance, a large module might split into two smaller ones as $\tG$ was
increased. With PISA, we were able to use a more direct approach. Once
we identified the modules, we computed the ``raw" ({\it i.e.}
pre-eliminations) condition-score vector
$\vect{r}$ for each module, and from these raw condition-score vectors,
we evaluated the condition correlations between modules (see Methods). 
Figure~\ref{Correlations} shows the condition correlations between 40 of
the
modules that we can put a name to. A large, positive correlation between
two modules can either indicate that the modules have many genes in
common, {\it e.g.} the genes of the arginine-biosynthesis module are
essentially a subset of the genes of the amino-acid-biosynthesis module, 
or, as in the toy model in 
Figs.~\ref{AbsorbedFixedPointFig}~and~\ref{PISAtoyex},
the modules have few/no genes in common, but the two
sets of genes are similarly regulated under many
conditions. In the toy model, the raw condition-score vectors
$\vect{r}_1$ and $\vect{r}_2$ correspond to the vectors in
Fig.~\ref{AbsorbedFixedPointFig}(a) and their correlation,
$\vect{r_1}\cdot\vect{r_2}/(\abs{\vect{r_1}}\abs{\vect{r_2}})$,
is simply the cosine of the angle between them.
A real example of this second type of correlation is provided by 
the ribosomal-protein module (104 genes) and the rRNA-processing module
(144 genes). They have no genes in
common, but the correlation between them is very high, 0.76.

\ignore{The ribosomal proteins module (136 genes) and the rRNA
processing module (102 genes) are a good example of the latter: They
have no genes in common, but the correlation between them is
0.73\ignore{Average correlation between genes is 0.52}.}

Out of the 6206 genes included in the expression data, 2626 genes appeared in at least one module, and 923 genes appeared in more than one module\footnote{We have adjusted for the facted that for some modules there are several versions that are very similar.}. No genes appeared in more than 4 different modules.

\ignore{\section{Medical Applications}

The module-finding algorithms discussed here, including SA/ISA and
PISA have potential applications in medicine. The ``conditions'' in
which gene-expression data is obtained can equally well represent
different tissues and/or different patients. Application of the
algorithms in these cases could help uncover ``disease modules'',
{\it i.e.} sets of genes coregulated in certain tissues 
under certain disease conditions. Knowledge of these modules could
prove valuable in diagnosis and treatment of the disease conditions.
For example, the gene-expression profile of a new patient could
be scanned for patterns of expression corresponding to previously
identified disease modules. This would directly aid in disease
diagnosis. Very accurate diagnosis obtained in this way could help 
guide treatment protocols, particularly if a database of diagnoses,
treatments, and outcomes for previous patients was available for
reference.
}

\section{Acknowledgements}

We wish to thank J. Ihmels and N. Barkai for sharing their data set,
and Rahul Kulkarni for valuable discussions.

C.T. acknowledges support from the National Key Basic Research Project
of China (No. 2003CB715900).

\ignore{This breaks the symmetry between genes and conditions, as
pointed out
in~\cite{Barkai2}, but in reality genes and conditions are not
symmetrical
at all: Each gene belongs to a fixed (and presumably fairly small)
set of transcription modules, while the conditions are whatever the
experiment design says - they could affect very few or very many
modules.
If the experiments from which the data is collected, generally affect
only
a few modules each, then thresholding on condition scores makes sense.
If, however, the experiments are designed to affect many modules
simultaneously, in different combinations (note that this gives the most
information from the least number of experiments, although that
information
is not easily extracted just by looking at the data), then thresholding
the
condition scores is not a good idea.}

\clearpage
\setcounter{figure}{0}
\setcounter{equation}{0}
\setcounter{section}{0}
\renewcommand{\thesection}{}
\renewcommand{\thefigure}{S\arabic{figure}}
\renewcommand{\theequation}{S\arabic{equation}}
\section{Supporting material}

\subsection{Normalization}

Here we review in detail the normalization procedure employed in PISA. The most obvious requirement for the normalization is that scores for different genes must be comparable. The procedure itself is as follows: Given a matrix $\mat{E}$ of log-ratio gene-expression data, we first set the average to zero for each condition,
\beq
  (\mat{E}^\prime)_{gc} = (\mat{E})_{gc}-\avg{(\mat{E})_{g^\prime c}}_{g^\prime},
  \label{RemoveConditionAverage}
\eeq
and then normalize to zero mean and unit variance for each gene, giving $\Eg$, which is used in PISA to calculate gene scores:
\beqar
  (\mat{E}^{\prime\prime})_{gc} & = & (\mat{E^\prime})_{gc}-\avg{(\mat{E^\prime})_{gc^\prime}}_{c^\prime} \label{RemoveGeneAverage} \\
  (\Eg)_{gc} & = & (\mat{E}^{\prime\prime})_{gc}/\sqrt{\avg{(\mat{E}^{\prime\prime})_{gc^\prime}^2}_{c^\prime}}.
  \label{NormalizeByGene}
\eeqar
For this normalization to be consistent through the iterations in mISA, the different condition scores must also be comparable. To get the initial value $\Eorti$ of the matrix used to calculate condition scores, we divide $\Eg$ by the rms value for each condition:
\beq
  (\Eorti)_{gc} = (\Eg)_{gc}/\sqrt{\avg{(\Eg)_{g^\prime c}^2}_{g^\prime}}.
  \label{NormalizeByCondition}
\eeq

Note that a simple approach would be to normalize for both genes and conditions simultaneously and thus use only a single set of data\footnote{If $\Eg=\Eorti$ initially, then it is equivalent to keep $\Eg$ constant or use $\Eg=\Eort$, which is updated every time PISA finds a module.}---this could be easily accomplished by alternately normalizing over conditions and genes a few times; the data converge quickly. There is, however, a risk of losing significant features of the data through excessive normalization. For some conditions, the typical change in expression levels may be very large, while for others it may be negligible, and it would be misleading to always normalize these to the same level; at the very least, this would give a lower signal to noise ratio. Therefore, we have chosen to normalize $\Eg$ over genes but not conditions, allowing conditions with large changes in expression level to make a proportionately larger contribution to gene scores. For genes, however, it is reasonable to always normalize to the same level. If two genes are in the same module, then there is little reason to consider the gene with the larger dynamical range to be more reliable than the other. That is why we use $\Eg$ to calculate $\Eorti$.

Also note a the difference between genes and conditions: The variance for a gene often depends on a small number of outlying values, and normalizing over genes prevents these from dominating. In contrast, the variance for a condition typically depends on many genes, and as such is a far more reliable quantity.

\ignore{
In~\cite{SA,ISA}, the authors used separate matrices for calculating gene scores and condition scores, due to the different requirements: When calculating gene scores, the data must be normalized, with zero mean and unit average, for each gene, in order for the scores for all the different genes to be comparable to the threshold. Similarly, the data used for calculating conditions scores must be normalized for each condition. A somewhat disturbing consequence of this approach is that some data points end up with different signs in the two matrices, {\it i.e.} the normalized data sets contradict each other! (Both values are typically very small in these cases.) One way to avoid this would be to normalize for both genes and conditions simultaneously and thus use only a single set of data---this is easily accomplished by alternately normalizing over conditions and genes a few times; the data converge quickly.
}

\subsection{Avoiding Positive Feedback}

The basic principle of SA, or an iteration of ISA/mISA, is to find the set of genes whose expression profiles most resemble those of the genes in the input set, either for all conditions (mISA) or for a selected subset of conditions (SA/ISA). Of course, the gene whose expression profile most resembles that of a given gene is the gene itself, thus
\ignore{Ignoring the difference between $\Eort$ and $\Eg$, the gene scores $(\vect{s}_i^{\rm G})_g$ for iteration $i$ in mISA are proportional to the weighted sums of the correlations between gene $g$ and all the genes in the input set $\vect{m}_i^{\rm G}$ for that iteration. Thus,}
there is a potential for significant positive feedback. Adding one gene to the input set would typically increase the score of that gene far more than the score of any other gene. As a consequence of positive feedback, adding one gene to the gene vector of a fixed point would have a considerable chance of yielding another fixed point, and a small set of genes could be a fixed point even if the genes were completely uncorrelated.

In PISA, we only find each module (or combination) once for each run, and it is important to be as certain as possible that we have the correct genes. We avoid positive feedback by using leave-one-out scoring for genes that had nonzero weight at the start of the iteration, {\it i.e.} we remove the contribution from gene $g$ from the condition scores $\vect{s}_i^{\rm{C}}$ before we use these scores to calculate the new score for gene $g$:

$$(\vect{s}_i^{\rm{G}})_g \equiv
\frac{(\Eg)_{g\_} 
(\vect{s}_i^{\rm{C}}-[\Eort\transp)_{\_g}(\vect{m}_i^{\rm{G}})_g]} 
{\abs{\vect{s}_i^{\rm{C}}-(\Eort\transp)_{\_g}(\vect{m}_i^{\rm{G}})_g}},$$
where $(\mat{A})_{j\_}$ is row $j$ of matrix $\mat{A}$, and $(\mat{A})_{\_j}$ is column $j$ of matrix $\mat{A}$. With a Gaussian distribution of the background noise, this approach is very close to neutral, {\it i.e.} adding a gene will neither affect that gene's score, nor will it significantly change $\sigma^{\rm 70\%}$ of the gene-score distribution.

Without positive feedback, fixed points may be marginally stable (or even unstable, {\it i.e.} a limit cycle), thus we do not require a true fixed point; we accept any gene vector reached after 20 iterations in mISA, as long as it contains at least 5 genes.

In SA/ISA, the authors do not eliminate positive feedback. Indeed it would be difficult to do so, as adding/removing a gene can change which conditions have scores exceeding the condition threshold. Apart from this complication, the feedback in SA/ISA is proportional to the number of conditions that make the threshold. For small modules, typically only a small fraction of the conditions have scores above the threshold, thus the feedback is lower than it would have been for PISA, which includes all conditions. For large modules, the feedback is only a minor effect in the first place. Nevertheless, the total number of fixed points for ISA is huge due to positive feedback---at a gene threshold coefficient $\tG=4.0$, there are, at a minimum, more than a million fixed points.

\subsection{Filters}

We chose the gene-score threshold as $7.0\sigma^{\rm 70\%}$ so that, on average, less than one gene would be included in a module purely due to background noise. This estimate assumed that the background noise had a Gaussian distribution. For most modules, the gene scores are the sums of contributions from many different conditions, and if these contributions are independent, as they should be for background noise, then the total background noise will have approximately a Gaussian distribution, regardless of the distribution for a single condition (central limit theorem). For modules that derive almost entirely from one or very few conditions, however, the distribution of gene scores may not be Gaussian.

\begin{figure}[htb]
\includegraphics[width=3.375in]{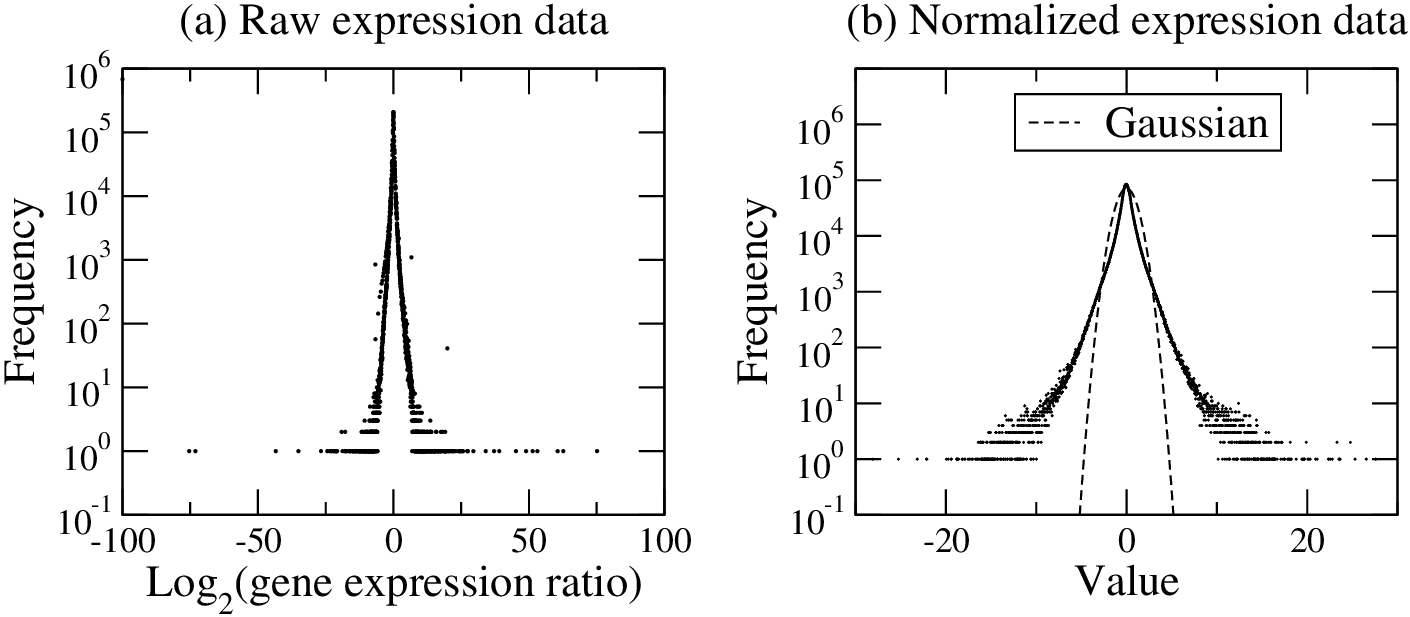}
\caption{Distributions of the yeast microarray data used
(6206 genes/ORFs, 1011 conditions). Roughly 10\% of the data was
invalid/missing
(not included in the distributions). The distribution is sharply cusped
and
has long tails, both before and after normalization (Eqs.~\ref{RemoveConditionAverage}--\ref{NormalizeByGene}).}
\label{distributions}
\end{figure}

While we do not know the true distribution of the background noise, it is reasonable to use the full distribution of the data as a worst case scenario. As shown in Fig.~\ref{distributions}, this distribution is far from Gaussian: it has a fairly sharp cusp at zero and long tails, even after normalization. For this distribution, more than 3.5\% of the values are outside the threshold $\pm 7.0\sigma^{\rm 70\%}$ (this is partially because the long tails contain many genes, and partially because $\sigma^{\rm 70\%}$ is small due to the sharp cusp), {\it i.e.} with a gene-expression matrix randomly drawn from this distribution, for any single condition one would expect to find a module with about 200 genes!

\begin{figure}[b]
\includegraphics[width=3.375in]{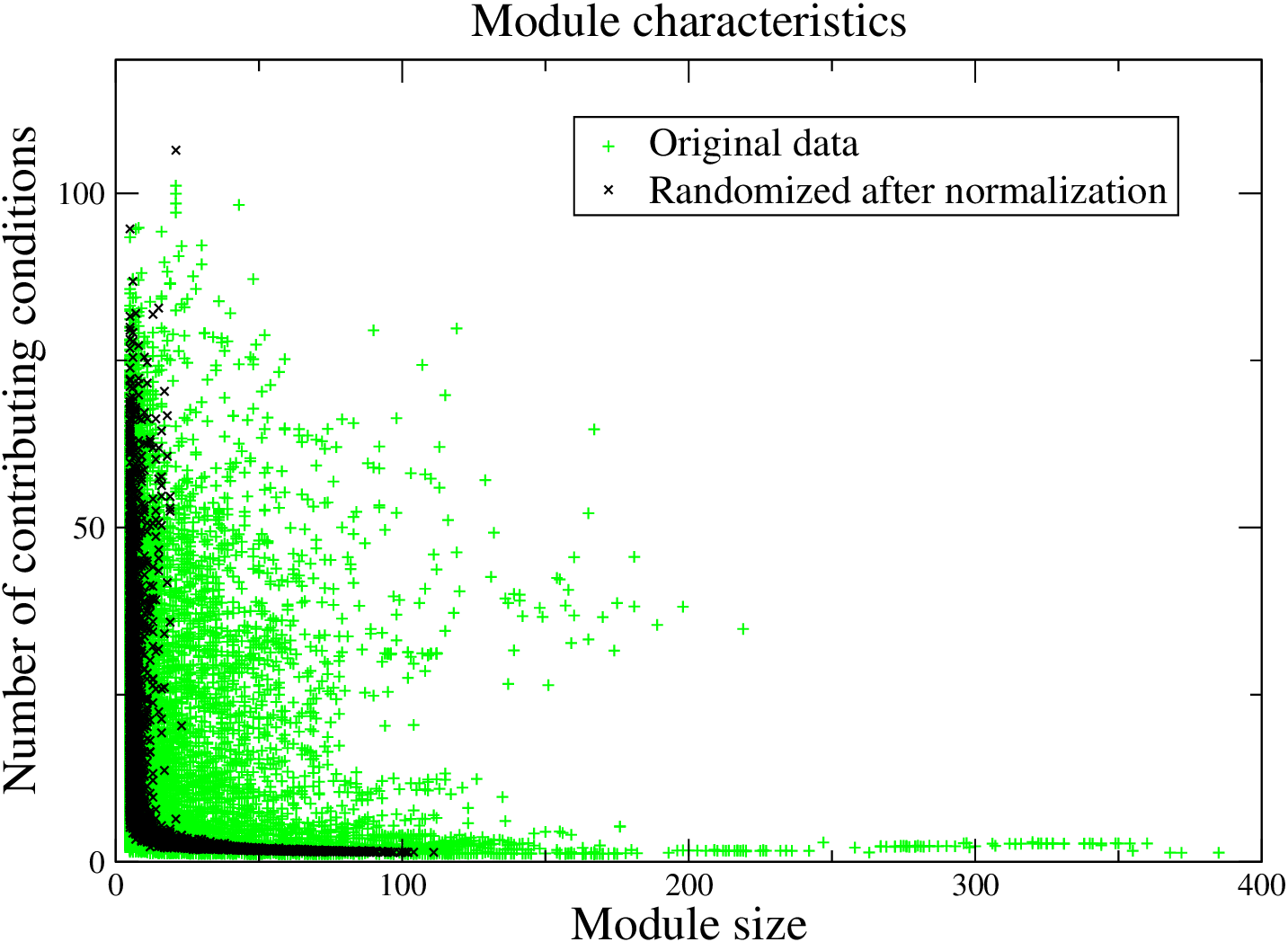}
\caption{The number of genes $n_{\Tm}^{\rm G}$ in a module $\Tm$ and the number of contributing conditions $n_{\Tm}^{\rm C}$ (see text) were two of the properties we used in our filters to eliminate false modules. PISA applied to a scrambled expression matrix (black) only yielded modules close to the axes (small $n_{\Tm}^{\rm G}$ or small $n_{\Tm}^{\rm C}$), while PISA run on the real data (green) yielded modules with both large $n_{\Tm}^{\rm G}$ and large $n_{\Tm}^{\rm C}$.}
\label{modulequality}
\end{figure}

We applied PISA to a matrix $\Eg$ that had been fully scrambled after normalization\footnote{Scrambling the matrix {\it after} normalization ensured that the distribution remained the same. The data were no longer exactly normalized for each gene, but the deviations were insignificant. Scrambling the data before normalization gave similar results.}. As shown in Fig.~\ref{modulequality}, PISA found many large modules that were based almost entirely on a single condition (however, as the modules were not based on {\it only} one condition, they were not as large as our estimate of 200, above), whereas modules based on many conditions were much smaller. We also applied PISA to a random matrix generated from a Gaussian distribution, and in that case PISA did not find any large modules (in 30 runs, PISA found 8 modules with 20 or more genes; the largest contained 26 genes). In both cases, the small modules found by PISA varied from run to run.

\comment{[This is an even more severe problem for ISA: As they use a condition-score threshold, it is possible to find modules that {\it really} depend on only a very few conditions. ISA ignores any module that does not have at least 5 (I think) conditions with at least 70\% (I think) of the score of the highest-scoring condition---this is (in my opinion) a somewhat more arbitrary way to implement a similar filter, and they don't say why they do this.]}

In order to eliminate these false modules we introduced a set of filters. For each preliminary module $\Tm$ we calculate the ``number of contributing conditions", given as $n_{\Tm}^{\rm C} = \sum_{c}(\vect{s}^{\rm C})_c^2/(\max\{(\vect{s}^{\rm C})_c\})^2$. We ignored any module for which the median of the numbers of contributing conditions for its preliminary modules was below 6 (this threshold worked well; it is somewhat above the threshold required to remove the false positives for the scrambled matrix). We also ignored all modules that had fewer than 5 genes or fewer than 5 contributing preliminary modules, and for modules with fewer than 10 genes we required that the ``consistency", defined as the average fraction of the genes in the preliminary modules that are in the full module, was above 0.55 (during post processing, we required that this fraction was above 0.2 for {\it each} preliminary module). These filters removed all but one of the modules found by PISA when applied to the scrambled matrix.

\ignore{
\subsection{More Results}

We here discuss the modules found by PISA in more detail. A total of 260 different modules passed our filters, however some of these were very similar to each other. We assigned each module a weight according to how much overlap there was with other modules that passed the filter, and using this weight, the modules we found roughly correspond to 143 unrelated modules. Figure~\ref{ModuleSizeDistribution} shows the distribution of modules sizes.

\begin{figure}[htb]
\includegraphics[width=3.375in]{ModuleSizeDistribution}
\caption{The distribution of sizes of modules found by PISA.}
\label{ModuleSizeDistribution}
\end{figure}

\comment{[HOW TO TAKE DATA?]}
}

\begin{table*}[h]
\begin{tabular}{|c|c|c|c|c|c|c|}
\hline
 & \# & \# & & Over.& Best & \\
Function & genes & cond. & Cons. & w/ISA & $t_{\rm G}$ & Freq.\\ \hline
Amino acid biosynthesis & 96 & 31.2 & 0.83 & 0.89 & 3.7 & 10090 \\
Arginine biosynthesis & 6 & 5.7 & 0.72 & 0.83 & 6.0 & 60 \\
Biotin synthesis \& transport & 6 & 6.5 & 0.80 & 0.67 & 5.5 & 7 \\
Lysine biosynthesis & 11 & 9.0 & 0.82 & 0.82 & 4.6 & 10 \\
De novo purine biosynthesis & 32 & 13.1 & 0.83 & 0.59 & 5.0 & 16 \\
Oxidative stress response & 69 & 23.8 & 0.91 & 0.32 & 3.4 & (1) \\
Aryl alcohol dehydrogenases & 6 & 15.4 & 0.62 & 0.83 & 4.9 & 8 \\
Proteolysis & 27 & 82.1 & 0.80 & 0.86 & 3.6 & 1661 \\
Trehalose \& hexose metabolism/conversion & 21 & 34.9 & 0.55 & 0.67 &
3.2 & 910 \\
COS genes & 11 & 9.2 & 0.49 & 1.00 & 3.3 & 756 \\
Heat shock & 52 & 42.8 & 0.78 & 0.38 & 3.2 & (1) \\
Repair of disulphide bonds & 26 & 41.6 & 0.73 & 0.58 & 3.5 & 15 \\
Calcium-calmodulin related & 41 & 32.5 & 0.78 & 0.73 & 3.0 & 2198 \\
Oxidative phosphorylation & 42 & 48.3 & 0.89 & 0.95 & 3.7 & 2600 \\
Gluconeogenesis, fatty acid beta-oxidation & 38 & 18.2 & 0.81 & 0.63 &
2.9 & 264\\
Mitochondrial ribosomal genes & 52 & 57.6 & 0.79 & 0.89 & 3.3 & 2291 \\
Transcription (RNA polymerase etc.)++ & 22 & 70.4 & 0.59 & 0.52 & 3.2 &
1 \\
Subtelomerically-encoded proteins & 36 & 48.2 & 0.94 & 1.00 & 3.9 & 6174
\\
Iron/copper uptake & 38 & 10.8 & 0.82 & 0.79 & 3.7 & 1704 \\
Coated vesicles/secretion & 25 & 47.6 & 0.61 & 0.64 & 3.7 & 4 \\
Phosphoglycerides biosynthesis & 33 & 36.1 & 0.86 & 0.61 & 2.9 & 27 \\
Hexose transporters & 10 & 33.9 & 0.74 & 0.60 & 3.8 & 41 \\
Galactose utilization & 23 & 17.4 & 0.84 & 0.74 & 3.2 & 686 \\
Mid sporulation & 97 & 11.7 & 0.90 & 0.70 & 2.7 & 6556 \\
Mating factors/receptors: a/$\alpha$ difference & 26 & 15.8 & 0.57 &
0.58 & 3.8 & 6\\
Mating & 110 & 31.1 & 0.89 & 0.75 & 2.7 & 24622 \\
Mating type a signaling genes & 6 & 18.6 & 0.26 & 0.83 & 5.5 & 22 \\
Mating genes for mating type a & 15 & 13.6 & 0.41 & 0.53 & 8.0 & 16 \\
Phosphate utilization & 27 & 24.4 & 0.89 & 0.81 & 3.3 & 5796 \\
Glycolysis & 19 & 26.9 & 0.54 & 0.89 & 3.7 & 91 \\
Ergosterol biosynthesis & 36 & 28.3 & 0.89 & 0.69 & 3.1 & 57 \\
Cell cycle G1/S & 66 & 39.1 & 0.80 & 0.81 & 3.7 & 4382 \\
Cell wall (bud emergence) & 17 & 42.7 & 0.76 & 0.94 & 4.0 & 63 \\
Cell cycle M/G1 & 35 & 31.4 & 0.82 & 0.89 & 3.9 & 952 \\
Cell cycle G2/M & 31 & 25.0 & 0.82 & 0.90 & 3.7 & 1258 \\
Uracil synthesis/permeases & 8 & 11.4 & 0.75 & 0.88 & 3.5 & 19 \\
Fatty acid synthesis++ & 22 & 49.4 & 0.86 & 0.50 & 3.1 & 2 \\
Histones & 19 & 34.6 & 0.67 & 0.53 & 3.4 & 2972 \\
Ribosomal proteins & 126 & 49.2 & 0.91 & 0.87 & 3.0 & 18661 \\
rRNA processing & 117 & 46.0 & 0.85 & 0.64 & 2.7 & 13355 \\ \hline
\end{tabular}
\caption{40 of the modules found by PISA that we could assign a name to.
For each module we list the number of genes in the module, the number of
conditions that had a significant contribution to the module, how
consistent
the module was from each run to the next, the maximal overlap with a
module
found by ISA (using 200,000 seeds at each threshold from 1.8 to 15.0),
the threshold value $t_{\rm G}$ at which that overlap was found, and how
many times such an ISA module was found.}
\end{table*}

\begin{figure}
\setlength{\unitlength}{0.6pt}
\fbox{
\begin{minipage}{3.200in}
\begin{flushleft}
\moduletitlesize
Module: Galactose induced genes
\modulesize

\vspace{10pt}

Number of genes: 23

Average number of contributing conditions: 18.1

Consistency: 0.84

Best ISA overlap: 0.74 at threshold 3.2, frequency 686
\genesize
\vspace{10pt}

\begin{picture}(300,100)(0,0)
\ignore{
      GAL10  30  651.086
       GAL7  30  638.409
       GAL1  30  632.368
       GAL3  30  492.239
       GAL2  30  424.969
    YPL066W  30  412.222
    YOR121C  30  370.898
      GAL80  30  362.546
      PCL10  30  350.308
       GCY1  30  339.059
       MLF3  30  306.93
    YDR010C  24  227.32
    YLR201C  22  200.349
       FUR4  21  188.24
       MUP3  18  178.833
     MRPL24  18  161.74
       OPT2  17  146.191
    YEL057C  16  135.229
}
\put(  65,  75){\gene{1}{GAL7}}
\put(   0,  75){\gene{1}{GAL10}}
\put( 130,  75){\gene{1}{GAL1}}
\put(  65,  25){\gene{4}{FUR4}}
\put( 195,  75){\gene{1}{GAL3}}
\put( 325,  50){\gene{0}{YDR010C}}
\put( 260,   0){\gene{2}{HXT3}}
\put( 325,  25){\gene{0}{YEL057C}}
\put( 130,  50){\gene{4}{PCL10}}
\put( 130,  25){\gene{4}{MUP3}}
\put( 130,   0){\gene{2}{HXT4}}
\put(   0,   0){\gene{2}{HXT1}}
\put( 195,   0){\gene{3}{HSL1}}
\put( 260,  75){\gene{1}{GAL2}}
\put(   0,  25){\gene{0}{YLR201C}}
\put(  65,  50){\gene{1}{GAL80}}
\put(  65,   0){\gene{2}{HXT2}}
\put( 195,  25){\gene{4}{MRPL24}}
\put( 260,  50){\gene{4}{MLF3}}
\put( 195,  50){\gene{1}{GCY1}}
\put(   0,  50){\gene{0}{YOR121C}}
\put( 325,  75){\gene{0}{YPL066W}}
\put( 260,  25){\gene{4}{OPT2}}
\ignore{
\put(  65,  75){\gene{1}{GAL7}}
\put(   0,  75){\gene{1}{GAL10}}
\put( 260,  75){\gene{1}{GAL1}}
\put( 195,  50){\gene{4}{FUR4}}
\put(   0,  50){\gene{1}{GAL3}}
\put( 325,  75){\gene{0}{YDR010C}}
\put( 260,   0){\gene{2}{HXT3}}
\put( 260,  50){\gene{0}{YEL057C}}
\put(  65,  25){\gene{4}{PCL10}}
\put( 325,  25){\gene{4}{MUP3}}
\put(  65,   0){\gene{2}{HXT4}}
\put( 195,   0){\gene{2}{HXT1}}
\put(   0,   0){\gene{3}{HSL1}}
\put( 130,  50){\gene{1}{GAL2}}
\put( 195,  25){\gene{0}{YLR201C}}
\put(   0,  25){\gene{1}{GAL80}}
\put( 130,   0){\gene{2}{HXT2}}
\put( 130,  25){\gene{4}{MRPL24}}
\put( 325,  50){\gene{4}{MLF3}}
\put( 260,  25){\gene{1}{GCY1}}
\put( 195,  75){\gene{0}{YOR121C}}
\put( 130,  75){\gene{0}{YPL066W}}
\put(  65,  50){\gene{4}{OPT2}}
}
\end{picture}

\vspace{20pt}
\gene{0}{0} \modulesize Unknown \genesize

\vspace{5pt}
\gene{1}{1} \modulesize Galactose induced genes \genesize

\vspace{5pt}
\gene{2}{2} \modulesize Hexose transporters (downregulated) \genesize

\vspace{5pt}
\gene{3}{3} \modulesize Other, downregulated \genesize

\vspace{5pt}
\gene{4}{4} \modulesize Other \genesize

\vspace{10pt}
\includegraphics[width=3.2in]{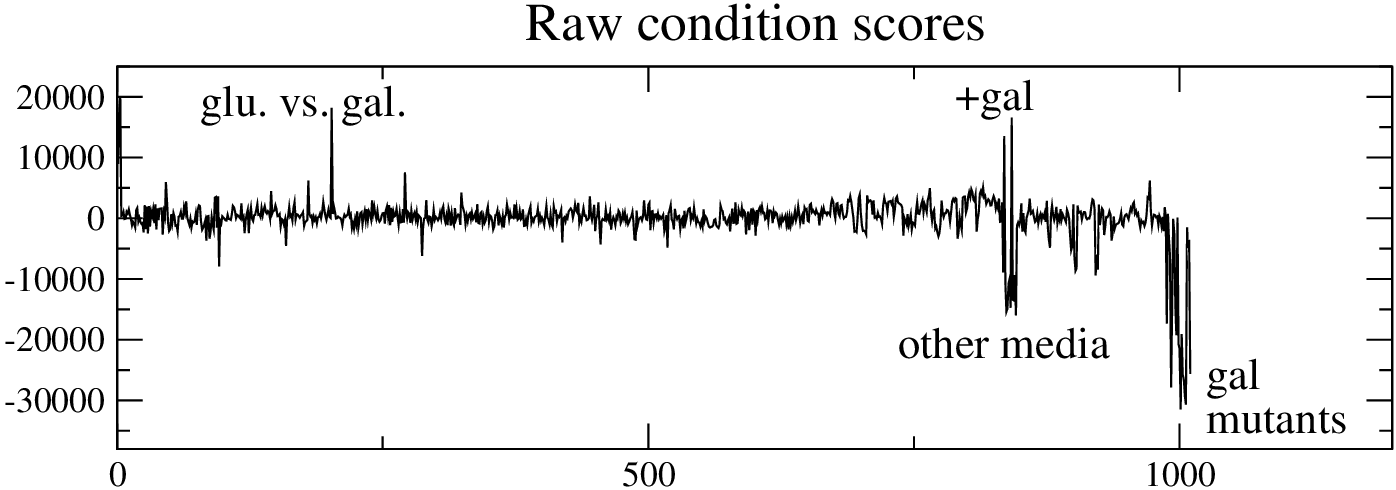}
\end{flushleft}
\end{minipage}
}
\caption{The galactose induced module found with PISA. This module turns
on
GAL genes and also, as a weaker effect, represses a number of hexose
transporters.}
\label{GalactoseModule}
\end{figure}

\begin{figure}
\setlength{\unitlength}{0.6pt}
\fbox{
\begin{minipage}{3.200in}
\begin{flushleft}
\moduletitlesize
Module: Hexose transporters
\modulesize

\vspace{10pt}

Number of genes: 10

Average number of contributing conditions: 33.7

Consistency: 0.74

Best ISA overlap: 0.6 at threshold 3.8, frequency 41
\genesize
\vspace{10pt}

\begin{picture}(300,50)(0,0)
\ignore{
       HXT3  28  292.002
       HXT4  28  260.904
       HXT2  27  231.856
       HXT6  26  228.681
       HXT7  25  192.247
       GAL2  25  182.687
    YKR075C  21  167.997
       HXT1  19  131.751
       MIG2  15  109.509
       HXT8  18  102.121
}
\put( 260,  25){\gene{1}{HXT7}}
\put( 195,  25){\gene{1}{HXT6}}
\put(   0,  25){\gene{1}{HXT3}}
\put( 130,   0){\gene{3}{MIG2}}
\put(  65,  25){\gene{1}{HXT4}}
\put(  65,   0){\gene{1}{HXT1}}
\put( 195,   0){\gene{1}{HXT8}}
\put(   0,   0){\gene{4}{YKR075C}}
\put( 325,  25){\gene{2}{GAL2}}
\put( 130,  25){\gene{1}{HXT2}}
\ignore{
\put( 130,  25){\gene{1}{HXT7}}
\put(   0,  25){\gene{1}{HXT6}}
\put(  65,  25){\gene{1}{HXT3}}
\put(   0,   0){\gene{3}{MIG2}}
\put( 195,  25){\gene{1}{HXT4}}
\put( 325,  25){\gene{1}{HXT1}}
\put( 195,   0){\gene{1}{HXT8}}
\put( 260,  25){\gene{4}{YKR075C}}
\put(  65,   0){\gene{2}{GAL2}}
\put( 130,   0){\gene{1}{HXT2}}
}
\end{picture}

\vspace{20pt}
\gene{1}{1} \modulesize Glucose transporter \genesize

\vspace{5pt}
\gene{2}{2} \modulesize Galactose/glucose transporter \genesize

\vspace{5pt}
\gene{3}{3} \modulesize Glucose suppression regulator \genesize

\vspace{5pt}
\gene{4}{4} \modulesize Similar to glucose suppression regulator
\genesize

\vspace{10pt}
\includegraphics[width=3.2in]{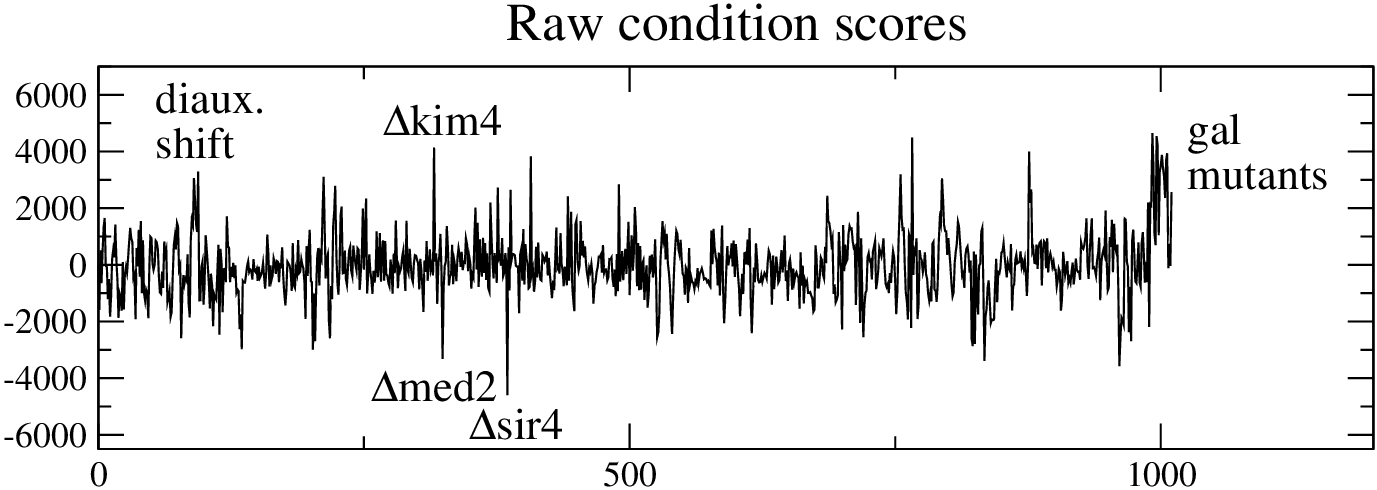}
\end{flushleft}
\end{minipage}
}
\caption{The hexose transporter module found with PISA. In this module
(which is consistently found after the galactose induced module), the
hexose transporter genes are co-regulated with GAL2, the galactose
permease,
whereas they were counter-regulated in the galactose induced module.
}
\label{HexoseModule}
\end{figure}

\begin{figure}
\setlength{\unitlength}{0.6pt}
\fbox{
\begin{minipage}{3.250in}
\begin{flushleft}
\moduletitlesize
Module: Peroxide shock
\modulesize

\vspace{10pt}

Number of genes: 69

Average number of contributing conditions: 23.9

Consistency: 0.91

Best ISA overlap: 0.34 at threshold 3.4, frequency (1)
\genesize
\vspace{10pt}
\begin{picture}(300,300)(0,0)
\ignore{
    YKL071W  27  440.999
       GPX2  27  439.203
    YCR102C  27  407.171
       FLR1  27  404.604
       AAD6  27  395.01
    YLR108C  27  393.525
    YDR132C  27  382.926
       GSH1  27  376.947
    YLR460C  27  359.603
       ISU2  27  352.729
    YML131W  27  351.048
    YFL057C  27  349.464
      AAD15  27  343.579
       AAD4  27  340.833
       GTT2  27  328.238
       FRE1  27  301.459
       ATR1  27  300.928
    YKR071C  27  299.813
       TRR1  27  296.053
      AAD14  27  293.316
       SFA1  27  287.805
    YNL134C  27  285.864
    YDR453C  27  284.169
       SDL1  27  280.931
    YOR225W  27  279.217
    YNL260C  27  278.871
       CCP1  27  272.347
    YOL150C  27  270.386
       YSR3  27  268.748
    YGL114W  27  260.416
    YNR074C  27  259.628
       AAD3  27  257.615
       ISA2  27  252.627
       OYE3  27  249.483
       MRS4  27  237.294
       YAP1  27  233.256
       CYT2  27  230.506
       ECM4  27  228.089
       TRX2  26  228.088
    YKL070W  27  224.577
      LYS20  26  217.823
       RIB3  26  214.765
       MMT1  25  209.659
      TAH18  25  209.175
    YMR318C  25  198.007
      TRS31  25  197.644
       GRE2  24  189.743
    YGR223C  24  188.606
       OYE2  23  179.3
    YGR011W  24  177.868
       SOD2  23  170.43
       TTR1  22  166.371
       KSS1  21  150.661
    YKL086W  19  144.809
    YOL029C  19  143.744
       KTR2  20  143.164
      NBP35  19  139.893
       CIN5  18  134.513
       SOD1  18  132.523
       AHP1  17  125.702
       NFU1  17  123.387
       LYS7  17  123.159
       TSA1  16  115.828
       ROX1  16  115.701
    YGR010W  16  113.785
     CDC123  15  107.395
    YHR199C  15  106.447
    YHR111W  14  101.71
    YPL202C  14  96.9853
}
\put( 195, 275){\gene{4}{FLR1}}
\put(  65, 275){\gene{1}{GPX2}}
\put( 130, 275){\gene{2}{YCR102C}}
\put(  65, 150){\gene{2}{AAD3}}
\put( 130, 200){\gene{2}{SFA1}}
\put( 260, 125){\gene{4}{LYS20}}
\put(  65, 225){\gene{2}{AAD4}}
\put(   0, 250){\gene{0}{YDR132C}}
\put(   0, 200){\gene{1}{TRR1}}
\put( 260, 200){\gene{1}{YDR453C}}
\put( 195, 100){\gene{4}{TRS31}}
\put( 325, 125){\gene{4}{RIB3}}
\put( 195,  75){\gene{1}{TTR1}}
\put( 260, 275){\gene{2}{AAD6}}
\put( 325, 250){\gene{2}{YFL057C}}
\put( 130,  50){\gene{4}{NBP35}}
\put( 325, 175){\gene{4}{YGL114W}}
\put( 260,  25){\gene{4}{YGR010W}}
\put(  65,  75){\gene{0}{YGR011W}}
\put( 260,  75){\gene{4}{KSS1}}
\put( 130, 125){\gene{1}{TRX2}}
\put( 325, 100){\gene{0}{YGR223C}}
\put( 130,  75){\gene{1}{SOD2}}
\put(  65,   0){\gene{4}{YHR111W}}
\put(   0,  75){\gene{2}{OYE2}}
\put(   0,   0){\gene{0}{YHR199C}}
\put( 325, 200){\gene{4}{SDL1}}
\put(  65, 250){\gene{4}{GSH1}}
\put( 260,  50){\gene{1}{SOD1}}
\put(   0,  25){\gene{0}{NFU1}}
\put( 195, 125){\gene{0}{YKL070W}}
\put(   0, 275){\gene{0}{YKL071W}}
\put( 325,  75){\gene{0}{YKL086W}}
\put(   0, 125){\gene{4}{CYT2}}
\put( 260, 150){\gene{3}{MRS4}}
\put( 260, 175){\gene{3}{YSR3}}
\put(  65,  50){\gene{4}{KTR2}}
\put( 130, 175){\gene{1}{CCP1}}
\put( 325, 225){\gene{0}{YKR071C}}
\put(  65, 125){\gene{4}{ECM4}}
\put( 130, 225){\gene{4}{GTT2}}
\put( 325, 275){\gene{0}{YLR108C}}
\put( 325,  50){\gene{1}{AHP1}}
\put( 195, 225){\gene{1}{FRE1}}
\put( 325,  25){\gene{0}{CDC123}}
\put( 130, 250){\gene{0}{YLR460C}}
\put( 325, 150){\gene{3}{YAP1}}
\put( 130,  25){\gene{1}{TSA1}}
\put( 260, 225){\gene{3}{ATR1}}
\put( 260, 250){\gene{1}{YML131W}}
\put(  65,  25){\gene{1}{LYS7}}
\put(   0, 100){\gene{4}{MMT1}}
\put( 130, 100){\gene{2}{YMR318C}}
\put( 195, 200){\gene{0}{YNL134C}}
\put(  65, 175){\gene{0}{YNL260C}}
\put(  65, 200){\gene{2}{AAD14}}
\put(   0, 150){\gene{1}{YNR074C}}
\put(   0,  50){\gene{0}{YOL029C}}
\put( 195, 175){\gene{0}{YOL150C}}
\put( 260, 100){\gene{3}{GRE2}}
\put(   0, 225){\gene{2}{AAD15}}
\put( 195,  50){\gene{3}{CIN5}}
\put(   0, 175){\gene{0}{YOR225W}}
\put( 195, 250){\gene{4}{ISU2}}
\put( 195, 150){\gene{2}{OYE3}}
\put( 130,   0){\gene{0}{YPL202C}}
\put(  65, 100){\gene{1}{TAH18}}
\put( 195,  25){\gene{3}{ROX1}}
\put( 130, 150){\gene{4}{ISA2}}
\ignore{
\put(  65, 275){\gene{4}{FLR1}}
\put(   0, 275){\gene{1}{GPX2}}
\put( 130, 275){\gene{2}{YCR102C}}
\put(  65, 250){\gene{2}{AAD3}}
\put( 195, 275){\gene{2}{SFA1}}
\put(   0, 250){\gene{4}{LYS20}}
\put(  65, 175){\gene{2}{AAD4}}
\put( 325, 275){\gene{0}{YDR132C}}
\put( 260, 250){\gene{1}{TRR1}}
\put(   0, 225){\gene{1}{YDR453C}}
\put(  65, 200){\gene{4}{TRS31}}
\put( 130, 200){\gene{4}{RIB3}}
\put( 325, 225){\gene{1}{TTR1}}
\put( 260, 275){\gene{2}{AAD6}}
\put( 325, 250){\gene{2}{YFL057C}}
\put( 260, 225){\gene{4}{NBP35}}
\put( 195, 175){\gene{4}{YGL114W}}
\put(   0, 200){\gene{4}{YGR010W}}
\put( 195, 225){\gene{0}{YGR011W}}
\put( 325, 200){\gene{4}{KSS1}}
\put( 195, 250){\gene{1}{TRX2}}
\put( 130, 250){\gene{0}{YGR223C}}
\put(  65, 225){\gene{1}{SOD2}}
\put( 325, 175){\gene{4}{YHR111W}}
\put( 130, 225){\gene{2}{OYE2}}
\put( 130, 175){\gene{0}{YHR199C}}
\put( 260, 200){\gene{4}{SDL1}}
\put(   0, 175){\gene{4}{GSH1}}
\put( 195, 200){\gene{1}{SOD1}}
\put( 260, 175){\gene{0}{NFU1}}
\put( 130, 125){\gene{0}{YKL070W}}
\put(   0, 150){\gene{0}{YKL071W}}
\put( 130,  50){\gene{0}{YKL086W}}
\put( 130,  25){\gene{4}{CYT2}}
\put(  65,  50){\gene{3}{MRS4}}
\put( 260,  75){\gene{3}{YSR3}}
\put( 260,  25){\gene{4}{KTR2}}
\put( 195,  50){\gene{1}{CCP1}}
\put( 195,  75){\gene{0}{YKR071C}}
\put(   0,  25){\gene{4}{ECM4}}
\put( 325,  75){\gene{4}{GTT2}}
\put( 325, 100){\gene{0}{YLR108C}}
\put( 325,  50){\gene{1}{AHP1}}
\put( 260, 100){\gene{1}{FRE1}}
\put(   0,  75){\gene{0}{CDC123}}
\put( 260, 125){\gene{0}{YLR460C}}
\put( 325, 125){\gene{3}{YAP1}}
\put( 325,  25){\gene{1}{TSA1}}
\put(  65, 100){\gene{3}{ATR1}}
\put( 195, 125){\gene{1}{YML131W}}
\put(  65,  75){\gene{1}{LYS7}}
\put( 130, 100){\gene{4}{MMT1}}
\put( 195, 100){\gene{2}{YMR318C}}
\put(   0, 100){\gene{0}{YNL134C}}
\put( 195, 150){\gene{0}{YNL260C}}
\put( 260, 150){\gene{2}{AAD14}}
\put(   0, 125){\gene{1}{YNR074C}}
\put( 325, 150){\gene{0}{YOL029C}}
\put( 130, 150){\gene{0}{YOL150C}}
\put(  65, 125){\gene{3}{GRE2}}
\put(  65, 150){\gene{2}{AAD15}}
\put(   0,   0){\gene{3}{CIN5}}
\put(   0,  50){\gene{0}{YOR225W}}
\put( 130,  75){\gene{4}{ISU2}}
\put( 130,   0){\gene{2}{OYE3}}
\put(  65,   0){\gene{0}{YPL202C}}
\put(  65,  25){\gene{1}{TAH18}}
\put( 195,  25){\gene{3}{ROX1}}
\put( 260,  50){\gene{4}{ISA2}}
}
\end{picture}

\vspace{20pt}
\gene{0}{0} \modulesize Unknown \genesize

\vspace{5pt}
\gene{1}{1} \modulesize Peroxidase, superoxide dismutase, reductase
\genesize

\vspace{5pt}
\gene{2}{2} \modulesize Dehydrogenase \genesize

\vspace{5pt}
\gene{3}{2} \modulesize Other stress related genes \genesize

\vspace{5pt}
\gene{4}{3} \modulesize Other \genesize

\vspace{10pt}
\includegraphics[width=3.2in]{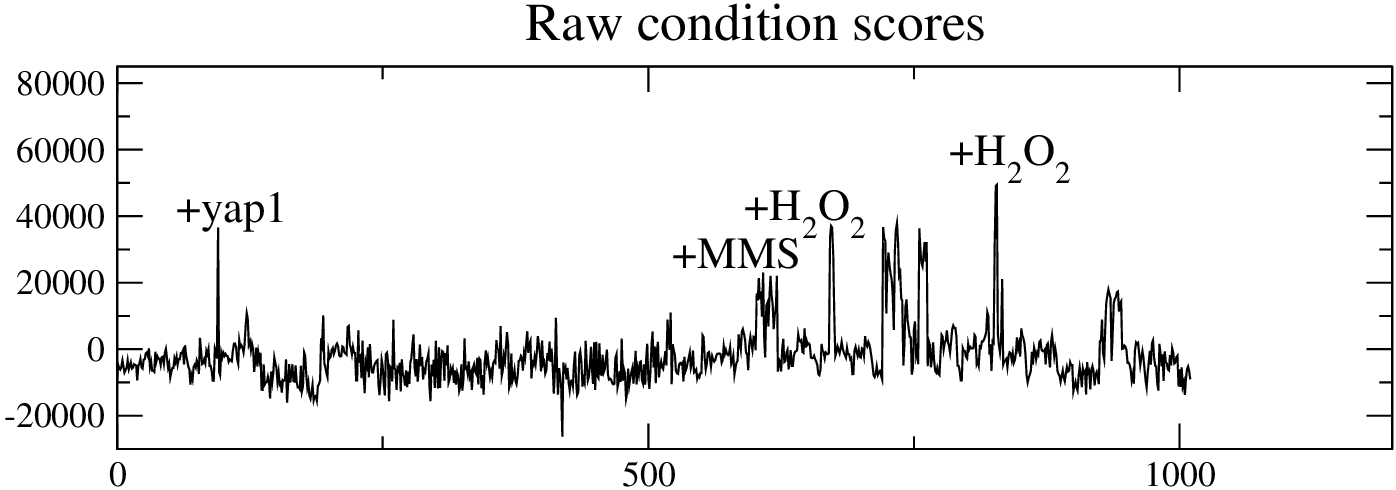}
\end{flushleft}
\end{minipage}
}
\caption{The oxidative stress response module found with PISA. This
module
is significantly more complete than the modules of comparable size found
by ISA.
}
\label{PeroxideShockModule}
\end{figure}

\ignore{
\begin{figure}
\setlength{\unitlength}{0.6pt}
\fbox{
\begin{minipage}{3.200in}
\begin{flushleft}
\moduletitlesize

Module: De novo purine biosynthesis
\modulesize

\vspace{10pt}
Number of genes: 29

Average number of contributing conditions: 13.797

Consistency: 0.859429

Best ISA overlap:  at threshold , frequency

\vspace{10pt}
\genesize
\begin{picture}(300,150)(0,0)
\put(  65,  50){\gene{2}{GCV3}}
\put( 260, 100){\gene{1}{ADE1}}
\put( 260,   0){\gene{3}{HIS7}}
\put( 325,  50){\gene{3}{HIS4}}
\put( 260,  25){\gene{4}{AIP2}}
\put(   0,  75){\gene{2}{GCV1}}
\put(  65,  25){\gene{0}{YDR089W}}
\put( 195,  50){\gene{1}{ADE8}}
\put( 195,  75){\gene{4}{CEM1}}
\put(   0,  25){\gene{2}{SER3}}
\put(  65,   0){\gene{2}{MET6}}
\put( 325,  75){\gene{1}{YGL186C}}
\put(  65,  75){\gene{1}{ADE5,7}}
\put( 260,  50){\gene{1}{ADE6}}
\put( 325,  25){\gene{1}{ADE3}}
\put( 130,   0){\gene{4}{ETF-BETA}}
\put( 130,  25){\gene{2}{SER2}}
\put(   0,   0){\gene{2}{SER33}}
\put(   0, 100){\gene{2}{MTD1}}
\put( 130, 100){\gene{2}{SHM2}}
\put( 325, 100){\gene{1}{ADE13}}
\put( 195,   0){\gene{4}{URA4}}
\put( 195, 100){\gene{1}{ADE17}}
\put(   0,  50){\gene{2}{GCV2}}
\put( 130,  75){\gene{1}{ADE4}}
\put( 130,  50){\gene{1}{ADE12}}
\put(  65, 100){\gene{1}{ADE2}}
\put( 260,  75){\gene{2}{SER1}}
\put( 195,  25){\gene{4}{YPR004C}}
\end{picture}

\vspace{20pt}
\gene{0}{0} \modulesize Unknown \genesize

\vspace{5pt}
\gene{1}{1} \modulesize Purine synthesis/transport \genesize

\vspace{5pt}
\gene{2}{2} \modulesize Tetrahydrofolate activation \genesize

\vspace{5pt}
\gene{3}{3} \modulesize Histidine biosynthesis \genesize

\vspace{5pt}
\gene{4}{4} \modulesize Other \genesize

\vspace{10pt}
\includegraphics[width=3.2in]{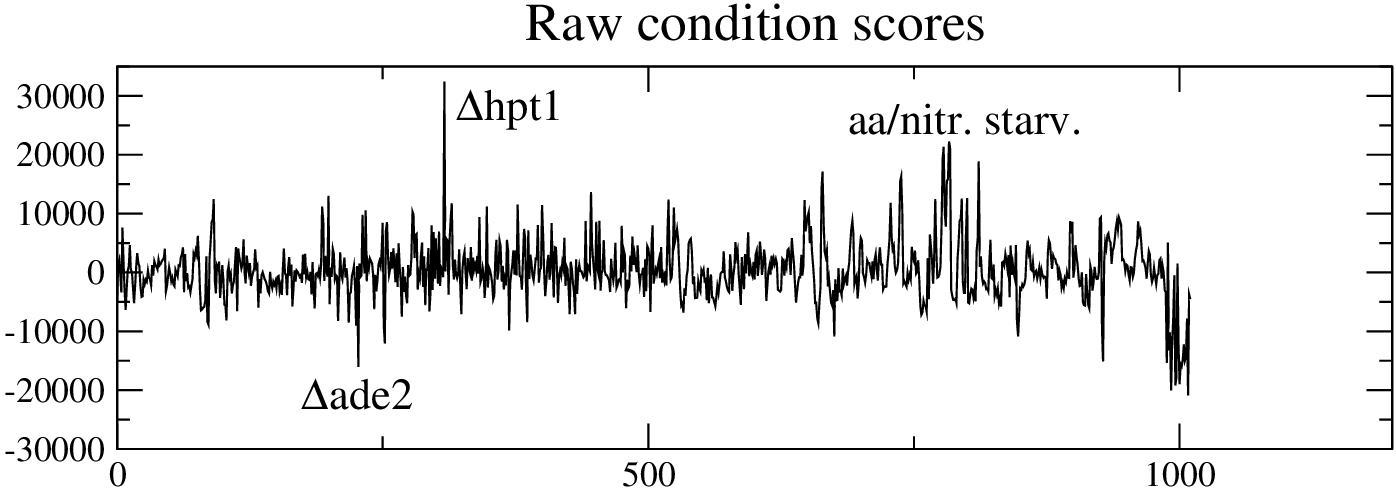}
\end{flushleft}
\end{minipage}
}
\caption{The de novo purine synthesis module found with PISA.}
\label{PurineModule}
\end{figure}

\begin{figure}
\setlength{\unitlength}{0.6pt}
\fbox{
\begin{minipage}{3.200in}
\begin{flushleft}
\moduletitlesize
Module: Galactose induced genes
\modulesize

\vspace{10pt}

Number of genes: 22

Average number of contributing conditions: 17.6144

Consistency: 0.863624

Best ISA overlap:  at threshold , frequency
\genesize
\vspace{10pt}

\begin{picture}(300,100)(0,0)
\put(  65,  75){\gene{1}{GAL7}}
\put(   0,  75){\gene{1}{GAL10}}
\put( 130,  75){\gene{1}{GAL1}}
\put( 195,  25){\gene{4}{FUR4}}
\put( 195,  75){\gene{1}{GAL3}}
\put(  65,  25){\gene{0}{YDR010C}}
\put( 195,   0){\gene{2}{HXT3}}
\put( 325,  25){\gene{0}{YEL057C}}
\put(  65,  50){\gene{4}{PCL10}}
\put( 130,  25){\gene{4}{MUP3}}
\put(  65,   0){\gene{2}{HXT4}}
\put( 130,   0){\gene{3}{HSL1}}
\put( 260,  75){\gene{1}{GAL2}}
\put(   0,  25){\gene{0}{YLR201C}}
\put( 130,  50){\gene{1}{GAL80}}
\put(   0,   0){\gene{2}{HXT2}}
\put( 325,  50){\gene{4}{MRPL24}}
\put( 260,  50){\gene{4}{MLF3}}
\put( 195,  50){\gene{1}{GCY1}}
\put(   0,  50){\gene{0}{YOR121C}}
\put( 325,  75){\gene{0}{YPL066W}}
\put( 260,  25){\gene{4}{OPT2}}
\end{picture}

\vspace{20pt}
\gene{0}{0} \modulesize Unknown \genesize

\vspace{5pt}
\gene{1}{1} \modulesize Galactose induced genes \genesize

\vspace{5pt}
\gene{2}{2} \modulesize Hexose transporters (downregulated) \genesize

\vspace{5pt}
\gene{3}{3} \modulesize Other, downregulated \genesize

\vspace{5pt}
\gene{4}{4} \modulesize Other \genesize

\vspace{10pt}
\includegraphics[width=3.2in]{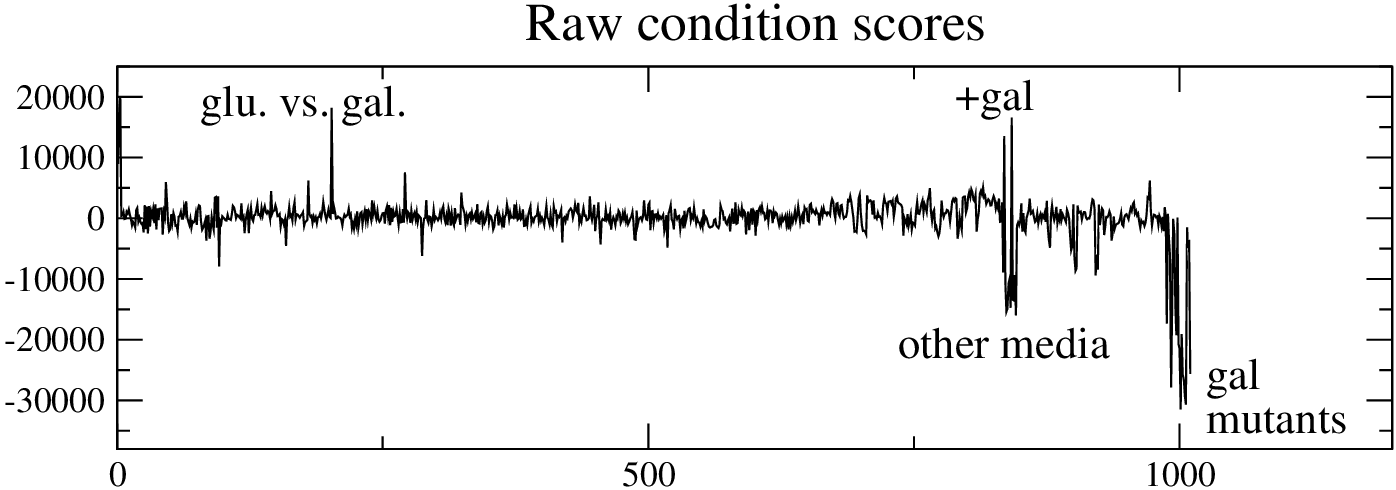}
\end{flushleft}
\end{minipage}
}
\caption{The galactose induced module found with PISA. This module turns
on
GAL genes and also, as a weaker effect, represses a number of hexose
transporters.}
\label{GalactoseModule}
\end{figure}

\begin{figure}
\setlength{\unitlength}{0.6pt}
\fbox{
\begin{minipage}{3.250in}
\begin{flushleft}
\moduletitlesize
Module: Peroxide shock
\modulesize

\vspace{10pt}

Number of genes: 56

Average number of contributing conditions: 25.3795

Consistency: 0.836174

Best ISA overlap:  at threshold , frequency
\genesize
\vspace{10pt}
\begin{picture}(300,300)(0,0)
\put( 260, 225){\gene{4}{FLR1}}
\put(  65, 225){\gene{1}{GPX2}}
\put( 130, 225){\gene{2}{YCR102C}}
\put(  65, 125){\gene{2}{AAD3}}
\put(   0, 150){\gene{2}{SFA1}}
\put( 260,  75){\gene{4}{LYS20}}
\put(  65, 175){\gene{2}{AAD4}}
\put(   0, 200){\gene{0}{YDR132C}}
\put( 260, 150){\gene{1}{TRR1}}
\put( 195, 150){\gene{1}{YDR453C}}
\put( 260,  50){\gene{4}{TRS31}}
\put(   0,  50){\gene{4}{RIB3}}
\put(   0,   0){\gene{1}{TTR1}}
\put( 195, 225){\gene{2}{AAD6}}
\put(   0, 175){\gene{2}{YFL057C}}
\put(  65, 100){\gene{4}{YGL114W}}
\put(  65,  25){\gene{0}{YGR011W}}
\put(   0,  25){\gene{4}{KSS1}}
\put(   0,  75){\gene{1}{TRX2}}
\put( 195,  50){\gene{0}{YGR223C}}
\put( 325,  25){\gene{1}{SOD2}}
\put( 260,  25){\gene{2}{OYE2}}
\put( 260, 125){\gene{4}{SDL1}}
\put(  65, 200){\gene{4}{GSH1}}
\put( 195,  75){\gene{0}{YKL070W}}
\put(   0, 225){\gene{0}{YKL071W}}
\put(  65,  75){\gene{4}{CYT2}}
\put( 325, 100){\gene{3}{MRS4}}
\put( 325, 125){\gene{3}{YSR3}}
\put(   0, 125){\gene{1}{CCP1}}
\put( 325, 150){\gene{0}{YKR071C}}
\put( 195, 100){\gene{4}{ECM4}}
\put( 130, 175){\gene{4}{GTT2}}
\put( 325, 225){\gene{0}{YLR108C}}
\put( 130, 150){\gene{1}{FRE1}}
\put( 130, 200){\gene{0}{YLR460C}}
\put( 130,  75){\gene{3}{YAP1}}
\put( 260, 175){\gene{3}{ATR1}}
\put( 325, 200){\gene{1}{YML131W}}
\put( 325,  75){\gene{4}{MMT1}}
\put( 325,  50){\gene{2}{YMR318C}}
\put(  65, 150){\gene{0}{YNL134C}}
\put( 195, 125){\gene{0}{YNL260C}}
\put( 195, 175){\gene{2}{AAD14}}
\put(   0, 100){\gene{1}{YNR074C}}
\put( 130,  25){\gene{0}{YOL029C}}
\put( 130, 125){\gene{0}{YOL150C}}
\put( 130,  50){\gene{3}{GRE2}}
\put( 260, 200){\gene{2}{AAD15}}
\put( 195,  25){\gene{3}{CIN5}}
\put( 325, 175){\gene{0}{YOR225W}}
\put( 195, 200){\gene{4}{ISU2}}
\put( 260, 100){\gene{2}{OYE3}}
\put(  65,  50){\gene{1}{TAH18}}
\put(  65,   0){\gene{3}{ROX1}}
\put( 130, 100){\gene{4}{ISA2}}
\end{picture}

\vspace{20pt}
\gene{0}{0} \modulesize Unknown \genesize

\vspace{5pt}
\gene{1}{1} \modulesize Peroxidase, superoxide dismutase, reductase
\genesize

\vspace{5pt}
\gene{2}{2} \modulesize Dehydrogenase \genesize

\vspace{5pt}
\gene{3}{2} \modulesize Other stress related genes \genesize

\vspace{5pt}
\gene{4}{3} \modulesize Other \genesize

\vspace{10pt}
\includegraphics[width=3.2in]{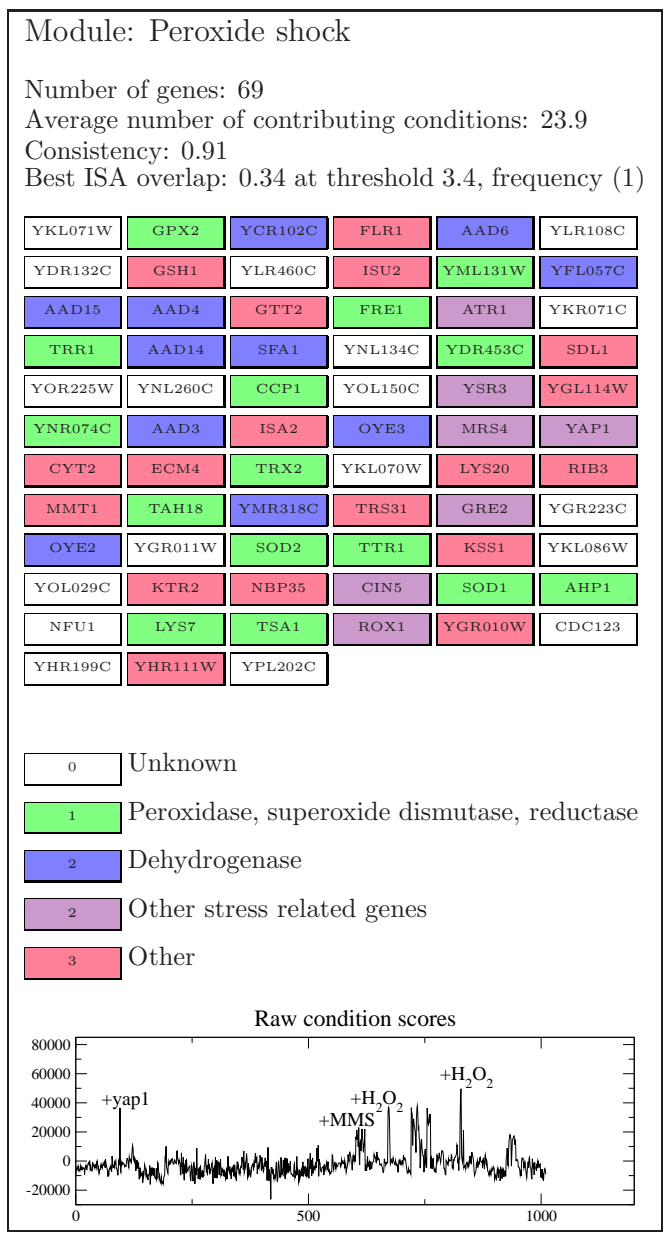}
\end{flushleft}
\end{minipage}
}
\caption{The oxidative stress response module found with PISA. This
module
is significantly more complete than the modules of comparable size found
by ISA.
}
\label{PeroxideShockModule}
\end{figure}

\begin{figure}
\setlength{\unitlength}{0.6pt}
\fbox{
\begin{minipage}{3.200in}
\begin{flushleft}
\moduletitlesize
Module: Hexose transporters
\modulesize

\vspace{10pt}

Number of genes: 12
Average number of contributing conditions: 33.1968
Consistency: 0.78495
Best ISA overlap:  at threshold , frequency
genesize
vspace{10pt}
\begin{picture}(300,50)(0,0)
put( 260,   0){\gene{3}{MTH1}}
put( 325,  25){\gene{1}{HXT7}}
put( 195,  25){\gene{1}{HXT6}}
put(   0,  25){\gene{1}{HXT3}}
put( 130,   0){\gene{3}{MIG2}}
put(  65,  25){\gene{1}{HXT4}}
put(  65,   0){\gene{1}{HXT1}}
put( 195,   0){\gene{1}{HXT8}}
put(   0,   0){\gene{3}{YKR075C}}
put( 260,  25){\gene{2}{GAL2}}
put( 130,  25){\gene{1}{HXT2}}
put( 325,   0){\gene{4}{YPR127W}}
end{picture}
\vspace{20pt}
gene{1}{1} \modulesize Glucose transporter \genesize
\vspace{5pt}
gene{2}{2} \modulesize Galactose/glucose transporter \genesize
\vspace{5pt}
gene{3}{3} \modulesize Glucose suppression regulator \genesize
\vspace{5pt}
gene{4}{4} \modulesize Other, downregulated \genesize
\vspace{10pt}
includegraphics[width=3.2in]{Hexose}
end{flushleft}
end{minipage}

caption{The hexose transporter module found with PISA. In this module
which is consistently found after the galactose induced module), the
hexose transporter genes are co-regulated with GAL2, the galactose
permease,
hereas they are counter-regulated in the galactose induced module. Many
of the genes that had a score slightly below the threshold are also
glucose suppression related.

label{HexoseModule}
end{figure}

begin{figure}
setlength{\unitlength}{0.6pt}
fbox{
begin{minipage}{3.200in}
begin{flushleft}
moduletitlesize
odule: Flocculation
modulesize
\vspace{10pt}
Number of genes: 6
Average number of contributing conditions: 15.5088
Consistency: 0.552039
Best ISA overlap:  at threshold , frequency
genesize
vspace{10pt}
\begin{picture}(300,25)(0,0)
ignore{
      FLO5  11  63.1642
      FLO9  11  57.3713
   YAL065C  11  52.6247
   YAR062W  11  50.6705
      FLO1   8  39.5147
   YHR213W   6  21.7364

put(  65,   0){\gene{2}{FLO9}}
put( 130,   0){\gene{2}{YAL065C}}
put( 260,   0){\gene{1}{FLO1}}
put( 195,   0){\gene{3}{YAR062W}}
put(   0,   0){\gene{1}{FLO5}}
put( 325,   0){\gene{3}{YHR213W}}
end{picture}

\vspace{20pt}
\gene{1}{1} \modulesize Flocculin \genesize

\vspace{5pt}
\gene{2}{2} \modulesize Similar to Flo1p \genesize

\vspace{5pt}
\gene{3}{3} \modulesize Similar to N-terminus of Flo1p \genesize

\vspace{10pt}
\includegraphics[width=3.2in]{Flocculation}
\end{flushleft}
\end{minipage}
}
\caption{The flocculation module found with PISA. 
\comment{There is some support for this in database A.
unknown genes for other modules
Use our modules to guide p-values of TF database.}
}
\label{FlocculationModule}
\end{figure}

}}}}
\begin{figure}
\setlength{\unitlength}{0.6pt}
\fbox{
\begin{minipage}{3.200in}
\begin{flushleft}
\moduletitlesize
Module: Zinc regulated genes
\modulesize

\vspace{10pt}

Number of genes: 8

Average number of contributing conditions: 29.0683

Consistency: 0.638515

Best ISA overlap: 0.88 at threshold 4.6, frequency 2
\genesize
\vspace{10pt}

\begin{picture}(300,50)(0,0)
\ignore{
       ZRT1  22  264.338
       ZRT3  22  202.015
       ZAP1  22  195.191
       ZRT2  18  128.968
    YOL154W  18  113.449
       INO1  17  109.054
       ADH4  19  108.413
    YNL254C  20  103.622
}
\put(   0,  25){\gene{1}{ZRT1}}
\put(   0,   0){\gene{4}{ADH4}}
\put( 130,  25){\gene{2}{ZAP1}}
\put( 325,  25){\gene{4}{INO1}}
\put(  65,  25){\gene{1}{ZRT3}}
\put( 195,  25){\gene{1}{ZRT2}}
\put(  65,   0){\gene{0}{YNL254C}}
\put( 260,  25){\gene{3}{YOL154W}}
\end{picture}

\vspace{20pt}
\gene{0}{0} \modulesize Unknown \genesize

\vspace{5pt}
\gene{1}{1} \modulesize Zinc transport/storage \genesize

\vspace{5pt}
\gene{2}{2} \modulesize Zinc-responsive transcription factor \genesize

\vspace{5pt}
\gene{3}{3} \modulesize Zinc metalloproteinase \genesize

\vspace{5pt}
\gene{4}{4} \modulesize Other \genesize

\vspace{10pt}
\includegraphics[width=3.2in]{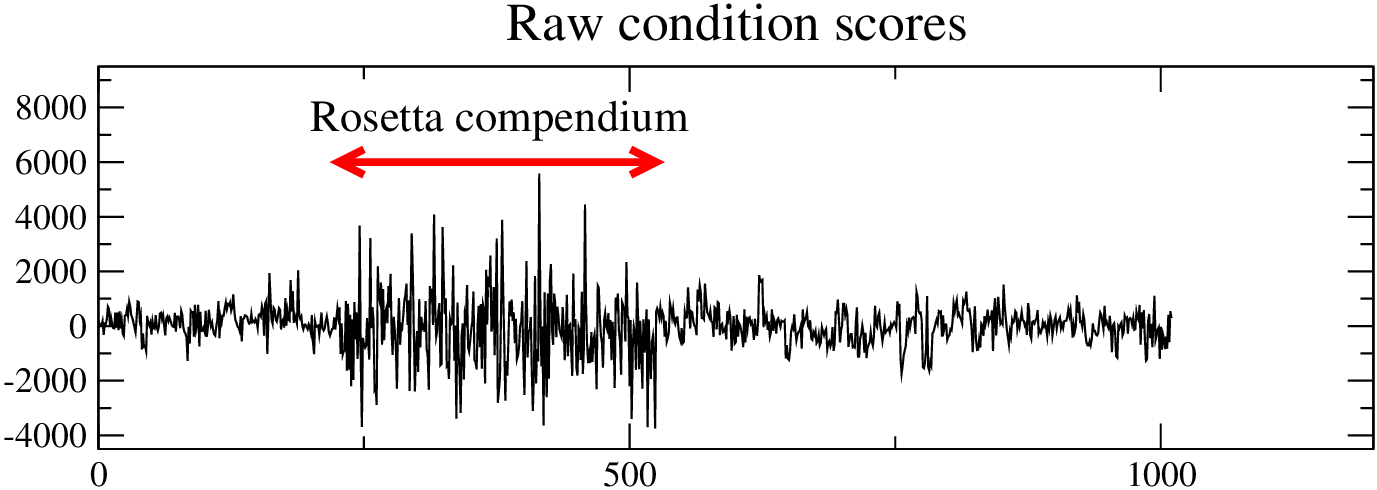}
\end{flushleft}
\end{minipage}
}
\caption{The zinc module found with PISA. This module has a high overlap
with
the group of genes bound by ZAP1 in database A (at $p$-value 0.001): The
ZRT1, ZRT2, ZRT3, ZAP1 and YNL254C genes make up 5 of the 6 lowest
$p$-values
(counting each pair of divergently transcribed genes only once), and the
remaining hits from database A (most with $p$-values above $10^{-4}$)
are likely to be mostly false positives. Based on this, it seems very
likely
than YNL254C, if functional, is regulated by and related to zinc. (ADH4
has also been shown to be zinc-regulated elsewhere.)
}
\label{ZincModule}
\end{figure}

\begin{figure}
\setlength{\unitlength}{0.6pt}
\fbox{
\begin{minipage}{3.200in}
\begin{flushleft}
\moduletitlesize
Module: Arginine regulation
\modulesize

\vspace{10pt}

Number of genes: 7

Average number of contributing conditions: 14.0667

Consistency: 0.548283

Best ISA overlap: 0.71 at threshold 6.0, frequency 60
\genesize
\vspace{10pt}

\begin{picture}(300,50)(0,0)
\put( 325,  25){\gene{1}{ARG5,6}}
\put(  65,  25){\gene{1}{ARG3}}
\put(   0,   0){\gene{2}{CAR2}}
\put( 260,  25){\gene{4}{CTF13}}
\put( 130,  25){\gene{1}{ARG1}}
\put(   0,  25){\gene{1}{ARG8}}
\put( 195,  25){\gene{1}{CPA1}}
\end{picture}

\vspace{20pt}
\gene{1}{1} \modulesize Arginine biosynthesis \genesize

\vspace{5pt}
\gene{2}{2} \modulesize Arginine degradation, downregulated \genesize

\vspace{5pt}
\gene{4}{4} \modulesize Other \genesize

\vspace{10pt}
\includegraphics[width=3.2in]{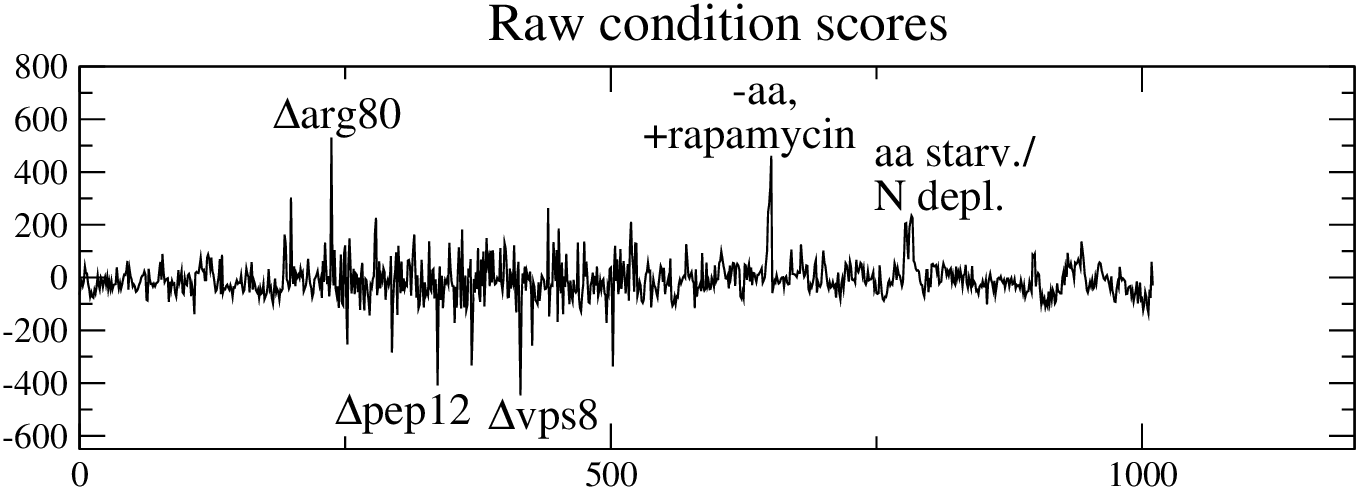}
\end{flushleft}
\end{minipage}
}
\caption{The arginine regulated module found with PISA. The module
agrees very well with what is known about regulation of arginine
metabolism
[F. Mesenguy and E. Dubois (2000) {\it Food tech. bio.} {\bf 38},
277-285]:
ARG1, ARG3, ARG5,6 and ARG8 are repressed by arginine through the
Arg80/Arg81/Mcm1 complex, while CAR2 (and CAR1, which is the 2nd highest
scoring gene that failed to make the module) is activated by the same
complex.
We also find CPA1, which is claimed to be regulated by arginine at the
translational
level---the mRNA is destabilized by a small peptide in the presence of
arginine. However, database A indicates that ARG1, ARG3, ARG5,6, ARG8
and CPA1 are
all bound by the Arg80/Arg81/Mcm1 complex.
}
\label{ArginineModule}
\end{figure}

\ignore{
\begin{figure}
\setlength{\unitlength}{0.6pt}
\fbox{
\begin{minipage}{3.200in}
\begin{flushleft}
\moduletitlesize
Module: IMP dehydrogenase
\modulesize

\vspace{10pt}

Number of genes: 5

Average number of contributing conditions: 17.6909

Consistency: 0.792111

Best ISA overlap:  at threshold , frequency
\genesize
\vspace{10pt}

\begin{picture}(300,25)(0,0)
\put(   0,   0){\gene{2}{IMD1}}
\put( 130,   0){\gene{2}{YAR075W}}
\put( 260,   0){\gene{3}{GLN3}}
\put(  65,   0){\gene{1}{IMD2}}
\put( 195,   0){\gene{2}{IMD3}}
\end{picture}

\vspace{20pt}
\vspace{5pt}
\gene{1}{1} \modulesize Inosine-5'-monophosphate dehydrogenase \genesize

\vspace{5pt}
\gene{2}{2} \modulesize Protein with similarity to Imd2p \genesize

\vspace{5pt}
\gene{3}{3} \modulesize Nitrogen regulatory protein, downregulated
\genesize

\vspace{10pt}
\includegraphics[width=3.2in]{IMD}
\end{flushleft}
\end{minipage}
}
\caption{The last IMD gene, IMD4, is the highest scoring gene not
included in
the module (but the score is far below that of GLN3).
}
\label{NewModule}
\end{figure}

\ignore{
\begin{figure}
\setlength{\unitlength}{0.6pt}
\fbox{
\begin{minipage}{3.200in}
\begin{flushleft}
\moduletitlesize
\begin{picture}(300,50)(0,0)
\end{picture}

\vspace{20pt}
\gene{0}{0} \modulesize Unknown \genesize

\vspace{5pt}
\gene{1}{1} \modulesize Zinc transport/storage \genesize

\vspace{5pt}
\gene{2}{2} \modulesize Zinc-responsive transcription factor \genesize

\vspace{5pt}
\gene{3}{3} \modulesize Zinc metalloproteinase \genesize

\vspace{5pt}
\gene{4}{4} \modulesize Other \genesize

\vspace{10pt}
\includegraphics[width=3.2in]{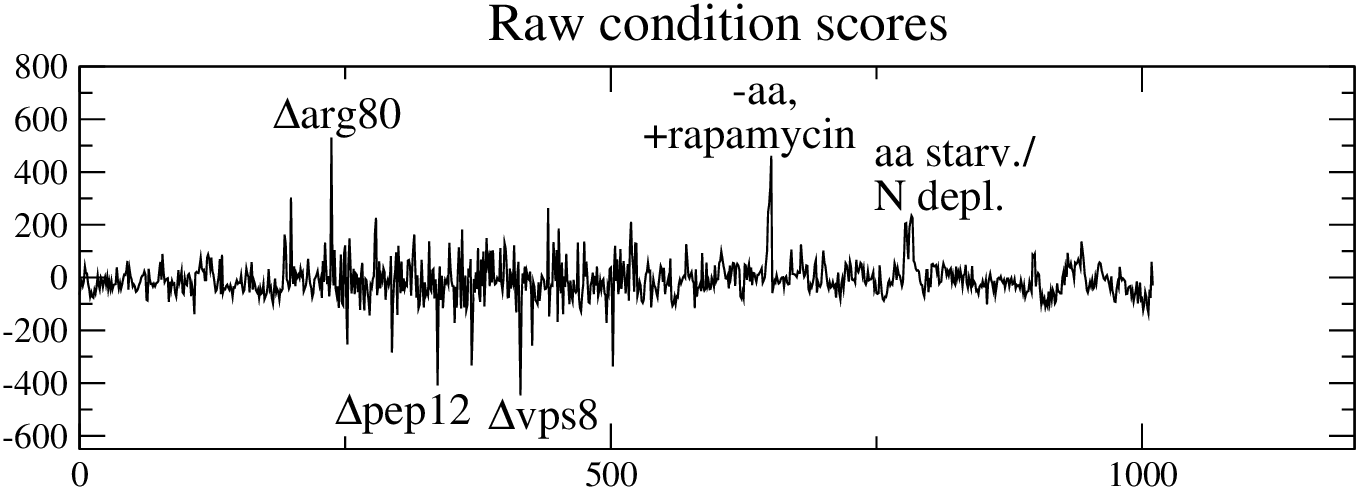}
\end{flushleft}
\end{minipage}
}
\caption{
}
\label{NewModule}
\end{figure}
}
}


\begin{thebibliography}{99}

\bibitem{Eisen_Cluster_analysis}
Eisen, M.B., Spellman, P.T., Brown, P.O. \& Botstein, D. (1998) {\it
Proc.
Natl. Acad. Sci.} {\bf 95}, 14863-14868.

\bibitem{Alon_Broad_patterns}
Alon, U., Barkai, N., Notterman, D.A., Gish, K., Ybarra, S., Mack, D. \&
Levine, A.J. (1999) {\it Proc. Natl. Acad. Sci.} {\bf 96}, 6745-6750.

\bibitem{Tamayo_Interpreting_patterns}
Tamayo, P., Slonim, D., Mesirov, J., Zhu, Q., Kitareewan, S.,
Dmitrovsky, E., Lander, E.S. \& Golub, T.R. (1999) {\it Proc. Natl.
Acad.
Sci.} {\bf 96}, 2907-2912.

\bibitem{Bittner_Data_analysis}
Bittner, M., Meltzer, P. \& Trent, J. (1999) {\it Nature Genet.} {\bf
22},
213-215.

\bibitem{Getz_Coupled_two-way}
Getz, G., Levine, E. \& Domany, E. (2000) {\it Proc. Natl. Acad. Sci.}
{\bf
97}, 12079-12084.

\bibitem{Califano_Analysis_of}
Califano, A., Stolovitzky, G. \& Tu, Y. (2000) {\it ISMB} {\bf 8},
75-85.

\bibitem{Cheng_Biclustering_of}
Cheng, Y. \& Church, G. (2000) {\it Proc. Int. Conf. Intell. Syst. Mol.
Biol.} {\bf 8}, 93-103.

\bibitem{Gasch_Exploring_the}
Gasch, A. \& Eisen, M.B. (2002) {\it Gen. Biol.} {\bf
3(11)}:research0059.1-0059.22.

\bibitem{Owen_Gene_Recommender}
Owen, A.B., Stuart, J., Mach, K., Villeneuve, A.M. \& Kim, S. (2003)
{\it Gen. Res.} {\bf 13}, 1828-1837.

\bibitem{Lazzeroni_Plaid_Models}
Lazzeroni, L. \& Owen, A. (2002) {\it Statistica Sinica}, {\bf 12(1)},
61-86.

\bibitem{SA}
Ihmels, J., Friedlander, G., Bergmann, S., Sarig, O., Ziv, Y. \& Barkai,
N. (2002)
{\it Nature Genet.} {\bf 31}, 370-377.

\bibitem{ISA}
Bergmann, S., Ihmels, J. \& Barkai, N. (2002) {\it Phys. Rev. E} {\bf
67}, 031902.

\bibitem{Lee_Transcriptional_Regulatory}
Lee, T. I. {\it et al.} (2002), {\it Science} {\bf 298}, 799-804.

\bibitem{Kellis_Sequencing_and}
Kellis, M., Patterson, N., Endrizzi, M., Birren, B. \& Lander, E. S.
(2003)
{\it Nature} {\bf 423}, 241-254.

\bibitem{Lyons_Genome-wide_characterization}
Lyons, T. J., Gasch, A. P., Gaither, L. A., Botstein, D., Brown, P. O.
\& Eide, D. J. (2000), {\it Proc. Natl. Acad. Sci.} {\bf 97}, 7957-7962.

\bibitem{Rosetta}
Hughes, T. R. {\it et al.} (2000) {\it Cell} {\bf 102}, 109-126.

\bibitem{GO}
The Gene Ontology Consortium (2001) {\it Genome Res.} {\bf 11},
1425-1433.

\bibitem{Messenguy_Regulation_Of}
Messenguy, F. \& Dubois, E. (2000) {\it Food tech. biotech.} {\bf 38}
277-285.

\end{thebibliography}
\end{document}